\documentclass[a4paper,11pt]{article}
\pdfoutput=1 

\usepackage{jinstpub} 

\usepackage[british]{babel}

\usepackage{subcaption}

\usepackage{siunitx,booktabs}
\newcommand{\ra}[1]{\renewcommand{\arraystretch}{#1}}
\usepackage{tabularx}
\usepackage{multirow}

\usepackage{url}
\usepackage{xspace}
\usepackage{color}
\usepackage[normalem]{ ulem } 

\newcommand{\catia}{\textsc{catia-ansys}\xspace}
\newcommand{\ebem}{{\textsc{el\-ec\-tro\-bem}}\xspace}
\newcommand{\garfield}{\textsc{garfield++}\xspace}
\newcommand{\gmsh}{\textsc{gmsh}\xspace}

\newcommand{\our}{{\textsc{ouroborosbem}}\xspace}

\newcommand{\bolsig}{\textsc{bolsig}\texttt{+}\xspace}

\title{OuroborosBEM: A fast multi-GPU microscopic Monte Carlo simulation for gaseous detectors and charged particle dynamics}

\author[1]{G.~Qu\'em\'ener\note{\url{https://orcid.org/0000-0001-6703-6655}},}
\author[2]{S.~Salvador\note{Corresponding author, \url{https://orcid.org/0000-0003-3444-7807}}}
\affiliation{Normandie Univ, ENSICAEN, UNICAEN, CNRS/IN2P3, LPC Caen, 14000 Caen, France }

\emailAdd{salvador@lpccaen.in2p3.fr}

\abstract{The design of gaseous detectors for accelerator, particle and nuclear physics requires simulations relying on multi-physics aspects. In fact, these simulations deal with the dynamics of a large number of charged particles interacting in a gaseous medium immersed in the electric field generated by a more or less complex assembly of electrodes and dielectric materials. We report here on a homemade software, called \our, able to tackle the different features involved in such simulations.
After solving the electrostatic problem for which a solver based on the boundary element method (BEM) has been implemented, particles are tracked and will microscopically interact with the gas medium. Dynamical effects have been included such as the electron-ion recombination process, the charging-up of the dielectric materials and other space charge effects that might alter the detector performances. These were made possible thanks to the nVidia CUDA language specifically optimised to run on Graphical Processor Units (GPUs) to minimize the computing times.
Comparisons of the results obtained for parallel plate avalanche counters and GEM detectors to literature data on swarm parameters fully validate the performances of \our. Moreover, we were able to precisely reproduce the measured gains of single and double GEM detectors as a function of the applied voltage.
}

\keywords{Simulation methods and programs, Accelerator modelling and simulations, Detector modelling and simulations II, Gaseous detectors}

\arxivnumber{2110.09214} 

\begin{document}
\bibliographystyle{unsrt}
\maketitle
\flushbottom

%
%
\section{Introduction}
\label{sec:Introduction}

Nowadays, simulating the behaviour of detectors is a mandatory phase of any particle and nuclear physics experiment especially for optimising and characterising its performances. When it comes to simulating gaseous detectors, not many tools are available. For instance \garfield~\citep{garfield}, an open-source software, has been used for years. Despite its ease of use and accurate macroscopic simulation of many gaseous detectors, the software suffers from the lack of specific detector behaviours. In fact, it cannot be used to generate complex electric fields and does not include any space charge or recombination effects. Moreover, the included microscopic simulation process where each electron is individually tracked and where interaction processes are generated following cross section data does not allow systematic studies due to an increased computation time. To overcome most of those limitations, we developed a microscopic Monte Carlo (MC) simulation software for Graphic Processor Units (GPUs) based on the nVidia CUDA programming language~\citep{cuda}.

Space charge effects may have a significant impact on the detector performances. They usually arise when approaching the operating limits of the system or when dealing with precision measurements. For instance, they might affect the gas gain in a large avalanche in Micro-Pattern Gaseous Detectors (MPGDs) such as Gaseous Electron Multipliers (GEMs) or MicroMegas. They are mostly attributed to high charge densities in the gas volume distorting the electric field and to charges collected by dielectric materials that are hardly evacuated to the nearest  electrode, also referred as the charging-up phenomenon. 

Other dynamic aspects encountered in gaseous detectors due to high charge densities are the charge recombination processes. Many processes of recombination can be considered~\citep{nasser} but they all relate to the association of species of opposing polarities forming a neutral entity that no longer participate in the generation of the measured electrical signal. Recombination is what mainly drives the non-linear response of air ionisation chambers in the case of high intensity beams monitoring~\citep{McManus:2020}. 

The simulation of those dynamical effects requires however a re-evaluation of the electric field at the beginning of every tracking step which makes them computationally challenging. In fact, each particle should be treated individually in position and velocity in order to affect the surrounding electrodes and neighbouring particles. This then relies on a microscopical approach rather than on a less accurate statistical one.

In this work, we will show how the computational power of GPUs made it possible to compute the electric field from complex geometries (with electrodes and dielectric materials) using a Boundary Element Method (BEM)~\cite{edurand} and including space charge effects and recombination processes at the microscopic level. We shall compare the results in terms of so-called swarm parameters such as drift velocities, diffusion coefficients or even first Townsend coefficients with literature data, the online Boltzmann equation solver \bolsig~\citep{bolsig} and an offline Python version of Magboltz, PyBoltz~\citep{ALATOUM}. We will also show some of the space charge effects on the electric field and potential distribution of some basic gaseous detectors.
%
%
\section{Electrostatic solver}
\label{sec:GeomSolver}

%
%
\subsection{Boundary integral equation formulation of the electrostatics problem}
\label{ssec:BIE}

\begin{figure}[hbt!]
\centering
\includegraphics[width=0.75\linewidth]{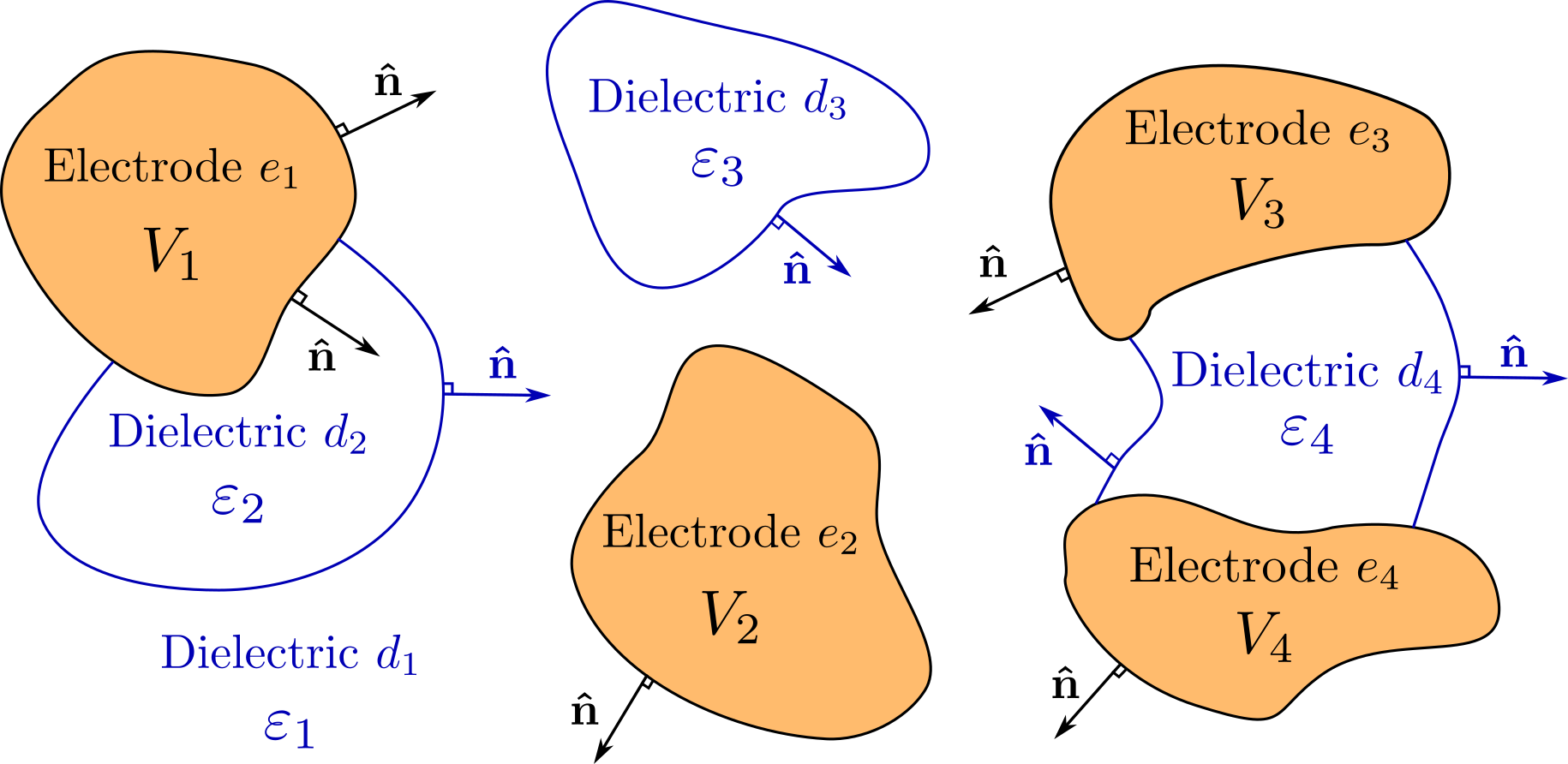}
\caption{Example of a composite system made of 4 conducting electrodes at known applied voltages $V_k$ ($k \in [1, 4]$) and 4 dielectric media  with absolute permittivities $\varepsilon_l$ ($l \in [1, 4]$). The electrode-to-dielectric interfaces ($ed$) are shown as black curves and the dielectric-to-dielectric interfaces ($dd$) are represented by blue curves. Also shown are the normal unit vectors $\mathbf{\hat{n}}$ at selected points on these different interfaces.}
\label{fig:composite}
\end{figure}

The knowledge of the electric scalar potential $V(\mathbf{r})$ and vector field $\mathbf{E}(\mathbf{r})$ at any location $\mathbf{r}$ within a gaseous detector is the central point for simulating this type of detector and investigating its performances. Let us consider a gaseous detector as a composite system of perfectly conducting electrodes at given applied voltages and of linear homogeneous isotropic dielectric materials (including vacuum) with given absolute permittivities as sketched in figure~\ref{fig:composite}. This composite system can be split into $N_{ed}$  electrode-to-dielectric interfaces and $N_{dd}$ dielectric-to-dielectric interfaces. In the rest of the text, the different quantities relative to  these interfaces or boundaries shall be labelled by the subscripts $ed$ and $dd$, respectively.

Determining the potential and field within such a system is equivalent to solve a boundary value problem with known boundary conditions. This type of  problem, usually formulated in terms of partial differential equations, is most often handled using the Finite Element Method (FEM) or the Finite Difference Method (FDM). In this work, the problem is reformulated in terms of Indirect Boundary Integral Equations (IBIE) satisfying Dirichlet and Neumann boundary conditions respectively on the $ed$- and $dd$-interfaces. In this indirect formulation, the unknowns are the equivalent surface charge densities $\sigma(\mathbf{r})$ on all the interfaces and are solved for by the boundary element method (BEM)~\cite{edurand}. Compared to FEM and FDM, BEM exhibits several advantages: it only requires to mesh the interfaces and not the whole volume, thus achieving similar precision with less computation time and memory requirements. This is especially true for geometries involving electrodes with large aspect ratios (e.g. an electrode with very small thickness for very large length and width) for which the 3D meshing quality would require special care. In addition, BEM does not require to interpolate or derive nodal solutions to obtain the potential and field at any location in space, the field components are therefore obtained with the same precision as the potential. Last but not least, BEM can easily deal with open systems as the boundary conditions are inherent to the IBIE formalism. 

The equivalent surface charge density on all the interfaces is the sum of a free surface charge density and a polarisation one:
\begin{equation}
  \sigma(\mathbf{r}) \, = \, \sigma_{\mathrm{F}}(\mathbf{r}) \, + \, \sigma_{\mathrm{P}}(\mathbf{r}).
\end{equation}
One should notice that $\sigma(\mathbf{r})$ reduces to $\sigma_{\mathrm{F}}(\mathbf{r})$ on the $ed$-interfaces and that, in the absence of free charges on the $dd$-interfaces, $\sigma(\mathbf{r})$ is equal to $\sigma_{\mathrm{P}}(\mathbf{r})$. The case of a non zero $\sigma_{\mathrm{F}}(\mathbf{r})$ on $dd$-interfaces is however very important to study the dielectric charging-up effect as for instance the collection of charged particles in a gaseous detector.

The potential $V(\mathbf{r})$ and field $\mathbf{E}(\mathbf{r})$ can be calculated at point $\mathbf{r}$ in space, once $\sigma(\mathbf{r})$ is known on every interfaces by the following equations:
\begin{eqnarray}
  V(\mathbf{r}) & = & \frac{1}{4\pi\varepsilon_0} \, \iint \limits_{\mathcal{S}} \frac{\sigma(\mathbf{r'})}{\| \mathbf{r} - \mathbf{r'} \|} \mathrm{d}^2\mathbf{r'}, \label{eq:BIEpot}\\
  \mathbf{E}(\mathbf{r}) & = & \frac{1}{4\pi\varepsilon_0} \, \iint \limits_{\mathcal{S}}  \sigma(\mathbf{r'}) \frac{\mathbf{r} - \mathbf{r'}}{\| \mathbf{r} - \mathbf{r'} \|^3} \mathrm{d}^2\mathbf{r'}, \label{eq:BIEfld}
\end{eqnarray}
where $\varepsilon_0$ is the vacuum permittivity and $\mathcal{S} = \mathcal{S}_{ed} \bigcup \mathcal{S}_{dd}$ is the total surface of the $ed$- and $dd$-interfaces. 
As mentioned previously, solving for $\sigma(\mathbf{r})$ requires to take into account some boundary conditions. The first condition is of Dirichlet type and states that on an $ed$-interface, e.g. on electrode $k$ whose surface is an iso-potential surface at known potential $V_k$, the surface charge density $\sigma(\mathbf{r})$ is linked to $V_k$ through the following equation (Fredholm equation of the first kind~\cite{eqfredholm}):
\begin{equation}
  V(\mathbf{r}) \, = \, V_k \, = \, \frac{1}{4\pi\varepsilon_0} \, \iint \limits_{\mathcal{S}} \frac{\sigma(\mathbf{r'})}{\| \mathbf{r} - \mathbf{r'} \|} \mathrm{d}^2\mathbf{r'} \qquad \forall \, \mathbf{r} \mbox{ on electrode } k. 
  \label{eq:DBC}
\end{equation}
The second condition is a Neumann boundary condition at a $dd$-interface between media 1 and 2 with respective absolute permittivities $\varepsilon_1 = \varepsilon_0 \varepsilon_{r,1}$ and $\varepsilon_2 = \varepsilon_0 \varepsilon_{r,2}$ (index $r$ indicates the relative permittivity). It relates the electric displacement vectors  $\mathbf{D}_1(\mathbf{r})$ and $\mathbf{D}_2(\mathbf{r})$ to a known free surface charge density $\sigma_{\mathrm{F}}(\mathbf{r})$ through the following relations:
\begin{eqnarray}
  \left[\mathbf{D}_2 (\mathbf{r})\, - \, \mathbf{D}_1(\mathbf{r})\right] \cdot \mathbf{\hat{n}}(\mathbf{r}) & = & \sigma_{\mathrm{F}}(\mathbf{r}), \label{eq:contdispl} \\
  \left[\varepsilon_2 \mathbf{E}_2(\mathbf{r}) \, - \, \varepsilon_1 \mathbf{E}_1(\mathbf{r})\right] \cdot \mathbf{\hat{n}}(\mathbf{r}) & = & \sigma_{\mathrm{F}}(\mathbf{r}), \label{eq:contfld} 
\end{eqnarray}
with $\mathbf{\hat{n}}(\mathbf{r})$ the normal unit vector pointing from media 1 to media 2 at point $\mathbf{r}$ on the $dd$-interface. Equation~\eqref{eq:contfld} implies that the normal component of the electric field is not continuous at a $dd$-interface even when $\sigma_{\mathrm{F}}$ is null.
Taking the limit of \eqref{eq:BIEfld} when the point $\mathbf{r}$ approaches $\mathcal{S}$ from medium 1 and from medium 2, the corresponding electric fields take the form:
\begin{eqnarray}
  \mathbf{E}_{1}(\mathbf{r}) & = & - \frac{\sigma(\mathbf{r})}{2\varepsilon_0} \mathbf{\hat{n}}(\mathbf{r}) \, + \, \frac{1}{4\pi\varepsilon_0} \, \iint \limits_{\mathcal{S}}  \sigma(\mathbf{r'}) \frac{\mathbf{r} - \mathbf{r'}}{\| \mathbf{r} - \mathbf{r'} \|^3} \mathrm{d}^2\mathbf{r'} \qquad \mbox{for } \mathbf{r} \longrightarrow \mathcal{S} \mbox{ from medium 1},
  \label{eq:fldinterf1} \\
  \mathbf{E}_{2}(\mathbf{r}) & = & + \frac{\sigma(\mathbf{r})}{2\varepsilon_0} \mathbf{\hat{n}}(\mathbf{r}) \, + \, \frac{1}{4\pi\varepsilon_0} \, \iint \limits_{\mathcal{S}}  \sigma(\mathbf{r'}) \frac{\mathbf{r} - \mathbf{r'}}{\| \mathbf{r} - \mathbf{r'} \|^3} \mathrm{d}^2\mathbf{r'} \qquad \mbox{for } \mathbf{r} \longrightarrow \mathcal{S} \mbox{ from medium 2}.
  \label{eq:fldinterf2}
\end{eqnarray}
Inserting \eqref{eq:fldinterf1} and \eqref{eq:fldinterf2} into \eqref{eq:contfld} leads to the following equation (Fredholm equation of the second kind~\cite{eqfredholm}):
\begin{equation}
  \sigma_{\mathrm{F}}(\mathbf{r}) \, = \, \frac{\varepsilon_2 \, + \, \varepsilon_1}{2\varepsilon_0} \sigma(\mathbf{r}) \, + \, \frac{\varepsilon_2 \, - \, \varepsilon_1}{4\pi\varepsilon_0} \, \iint \limits_{\mathcal{S}}  \sigma(\mathbf{r'}) \frac{(\mathbf{r} - \mathbf{r'}) \cdot \mathbf{\hat{n}}(\mathbf{r})}{\| \mathbf{r} - \mathbf{r'} \|^3} \, \mathrm{d}^2\mathbf{r'} \qquad \forall \, \mathbf{r} \mbox{ on } \mathcal{S}_{dd}.
  \label{eq:NBC}
\end{equation}

Equations~\eqref{eq:DBC} and \eqref{eq:NBC} form a set of boundary integral equations which relates the known quantities $V_k \,\, (k \in [1, N_{ed}])$ and $\sigma_{\mathrm{F},l}(\mathbf{r}) \,\, (l \in [1, N_{dd}]$) to the unknown $\sigma(\mathbf{r})$ on  $ed$- and $dd$-interfaces. To solve this set of equations for the surface charge distributions, we shall discretise the problem as shown in the following section.

%
%
\subsection{Discretisation of the boundary integral equations: from IBIE to BEM}
\label{ssec:Discretisation}

The discretisation process of the boundary integral equations is based on the  meshing of the geometry.
As mentioned in previous section, any set-up is described by $N_{ed}$ electrode-to-dielectric interfaces (corresponding to the surface of the electrodes) and $N_{dd}$ dielectric-to-dielectric interfaces between two dielectric media. The surface of electrode $e \,(e \in [1, N_{ed}])$ is meshed in $n_{e}$ flat polygonal cells (triangles, quadrangles, etc.) while the $dd$-interface $d \, (d \in [1, N_{dd}])$ is meshed in $n_d$ cells.
The full set-up is therefore represented by the set of cells $\mathcal{C} = \left\{ \mathcal{C}_i: i \in [1, \mathcal{N}] \right\}$ where
$\mathcal{N} \, = \, \sum_{e=1}^{N_{ed}} n_{e} \, + \, \sum_{d=1}^{N_{dd}} n_{d} \triangleq \mathcal{N}_{ed} \, + \, \mathcal{N}_{dd}$ is the total number of cells, each with an a priori unknown associated equivalent surface charge density $\sigma_i$ assumed constant over the 
cell surface.
The $n_e$ cells of electrode $e$ obviously share the same potential applied on  the whole electrode. The $n_d$ cells of $dd$-interface $d$ share the same permittivities $\varepsilon_1$ on the negative side and $\varepsilon_2$ on the outward-pointing normal (positive) side.

The known potential $V_i$ ($i \in [1, \mathcal{N}_{ed}]$) of cell $\mathcal{C}_i$ centred at position $\mathbf{r}_i$ is related to the 
unknown equivalent surface charge densities $\sigma_j$ of cells $\mathcal{C}_j$ ($j \in [1, \mathcal{N}]$) through the superposition 
principle as:
\begin{equation}
  V_i \, = \, \sum_{j=1}^{\mathcal{N}} \frac{\sigma_j}{4\pi\varepsilon_0} \, \iint \limits_{\mathcal{C}_j} \frac{1}{\| \mathbf{r}_i - \mathbf{r'} \|} \mathrm{d}^2\mathbf{r'} \, = \, \sum_{j=1}^{\mathcal{N}} \, Q_{ij} \, \sigma_j \qquad i \in [1, \mathcal{N}_{ed}],\label{eq:BEMpot}
\end{equation}
where we have introduced the matrix element $Q_{ij}$ for simplification and later use.
As $\sigma_j$ is assumed constant over the whole surface of cell $\mathcal{C}_j$, it has been moved outside of the surface integral.
Similarly, the known free surface charge density $\sigma_{\mathrm{F},i}$ ($i \in [\mathcal{N}_{ed}+1, \mathcal{N}]$) of cell $\mathcal{C}_i$ 
with normal unit vector $\mathbf{\hat{n}}_i$ is related to the unknown $\sigma_j$ through:
\begin{equation}
  \sigma_{\mathrm{F},i} \, =  \frac{\varepsilon_2 \, + \, \varepsilon_1}{2\varepsilon_0} \sigma_j \delta_{ij} \, + \, \frac{\varepsilon_2 \, - \, \varepsilon_1}{4\pi\varepsilon_0} \, \sum_{\substack{j=1 \\ i \neq j}}^{\mathcal{N}} \sigma_j \iint \limits_{\mathcal{C}_j}  \frac{(\mathbf{r}_i - \mathbf{r'}) \cdot \mathbf{\hat{n}}_i}{\| \mathbf{r}_i - \mathbf{r'} \|^3} \, \mathrm{d}^2\mathbf{r'}  \, = \, \sum_{j=1}^{\mathcal{N}} \, W_{ij} \, \sigma_j \qquad i \in [\mathcal{N}_{ed}+1, \mathcal{N}],\label{eq:BEMfld}
\end{equation}
where $\delta_{ij}$ is the usual Kronecker symbol and the matrix element $W_{ij}$ has been introduced for the same reasons as $Q_{ij}$.
In~\eqref{eq:BEMpot} and \eqref{eq:BEMfld}, the integrals over the surface of cell $\mathcal{C}_j$ only depend on $\mathcal{C}_j$ shape and on the relative position of the target point $\mathbf{r}_i$  with respect to $\mathcal{C}_j$ location. Analytic formulae have been derived for the surface integrals in \eqref{eq:BEMpot} and \eqref{eq:BEMfld}. 
These formulae are too complicated to be discussed here and shall be published in a separate article, however
it should be emphasised that special care has been taken to suppress numerical instabilities in the evaluation of these nearly singular integrals especially when $i = j$ where numerical divergences may occur.  Equation~\eqref{eq:BEMpot} (resp. \eqref{eq:BEMfld}) can be written for each $ed$-cell (resp. $dd$-cell) $\mathcal{C}_i$, leading to the following set of $\mathcal{N}$ linear algebraic equations with $\mathcal{N}$ unknowns:
\begin{equation}
   \begin{pmatrix} 
    V_1\\
    \vdots\\
    V_{\mathcal{N}_{ed}} \\
    \sigma_{\mathrm{F},\mathcal{N}_{ed}+1}\\
     \vdots\\
    \sigma_{\mathrm{F},\mathcal{N}_{ed}+\mathcal{N}_{dd}}\\
   \end{pmatrix}
    =
    \begin{pmatrix}
    Q_{1,1} & Q_{1,2}& \cdots  & Q_{1,\mathcal{N}}\\
    \vdots & \vdots & \ddots &\vdots \\ 
    Q_{\mathcal{N}_{ed},1} & Q_{\mathcal{N}_{ed},2}& \cdots  & Q_{\mathcal{N}_{ed},\mathcal{N}}\\
    W_{\mathcal{N}_{ed}+1,1} & W_{\mathcal{N}_{ed}+1,2}& \cdots  & W_{\mathcal{N}_{ed}+1,\mathcal{N}}\\
    \vdots & \vdots & \ddots &\vdots \\ 
     W_{\mathcal{N}_{ed}+\mathcal{N}_{dd},1} & W_{\mathcal{N}_{ed}+\mathcal{N}_{dd},2}& \cdots  & W_{\mathcal{N}_{ed}+\mathcal{N}_{dd},\mathcal{N}_{ed}+\mathcal{N}_{dd}}\\
   \end{pmatrix}
    \begin{pmatrix}
    \sigma_1\\
   \vdots\\
    \sigma_{\mathcal{N}_{ed}}\\
    \sigma_{\mathcal{N}_{ed}+1}\\
    \vdots\\
    \sigma_{\mathcal{N}_{ed}+\mathcal{N}_{dd}}\\
   \end{pmatrix}.
    \label{eq:matrixBEM}
\end{equation}
The matrix elements $Q_{ij}$ and $W_{ij}$ are computed for a given composite assembly/geometry leading to a dense square 
matrix on the contrary to FEM and FDM which involve sparse matrices of very large dimensions. The elements of this dense matrix are calculated on the GPU and the whole matrix is inverted using Basic Linear Algebra Subprograms (BLAS)~\cite{blackford2002updated} adapted for GPUs in the cuBLAS libraries~\cite{cuda}.

Having the inverted matrix at our disposal, makes it for example easy to change the applied voltages and/or free charge densities to obtain new $\sigma_i$.
Once the charge densities $\sigma_i$ have been determined on every cells, both the potential and field components can be evaluated at any location $\mathbf{r}$ in space without any interpolation: 
\begin{eqnarray}
  V(\mathbf{r}) & = & \sum_{i=1}^{\mathcal{N}} \, \frac{\sigma_i}{4\pi\varepsilon_0} \, \iint \limits_{\mathcal{C}_i} \frac{1}{\| \mathbf{r} - \mathbf{r'} \|} \mathrm{d}^2\mathbf{r'},  \label{eq:BEM2} \\
  \mathbf{E}(\mathbf{r}) & = & \sum_{i=1}^{\mathcal{N}} \, \frac{\sigma_i}{4\pi\varepsilon_0} \, \iint \limits_{\mathcal{C}_i} \frac{\mathbf{r} - \mathbf{r'}}{\| \mathbf{r} - \mathbf{r'} \|^3} \mathrm{d}^2\mathbf{r'}. \label{eq:BEM3}
\end{eqnarray}
In the same way as in \eqref{eq:BEMpot} and \eqref{eq:BEMfld}, integrals in \eqref{eq:BEM2} and \eqref{eq:BEM3} are analytically evaluated over the surface of each cell.

%
%
\subsection{Geometrical description of the set-up}
\label{ssec:GeomMod}

To compute the electric potential and field, we have adapted to the GPU a home made C/C\texttt{++} Poisson's solver based on BEM, \ebem \cite{Benali:2020}, that was developed ten to fifteen years ago.
This new GPU program is called \our\footnote{Available under GNU license at \url{https://gitlab.in2p3.fr/ouroboros/ouroboros_bem.git.}}.

In \our, a given set-up is implemented by first modelling the geometry of its compounding volumes and their corresponding electric properties such as applied voltages for the electrodes and the relative permittivities of the dielectric materials along with the gas species if the volume contains some gas. The knowledge of this geometry is not only necessary for the field computation but also for the particle dynamics to handle the loss of particles encountering a collision with neighbouring surfaces. 
\begin{figure}[htbp!]
\centering
\includegraphics[width=0.9\linewidth]{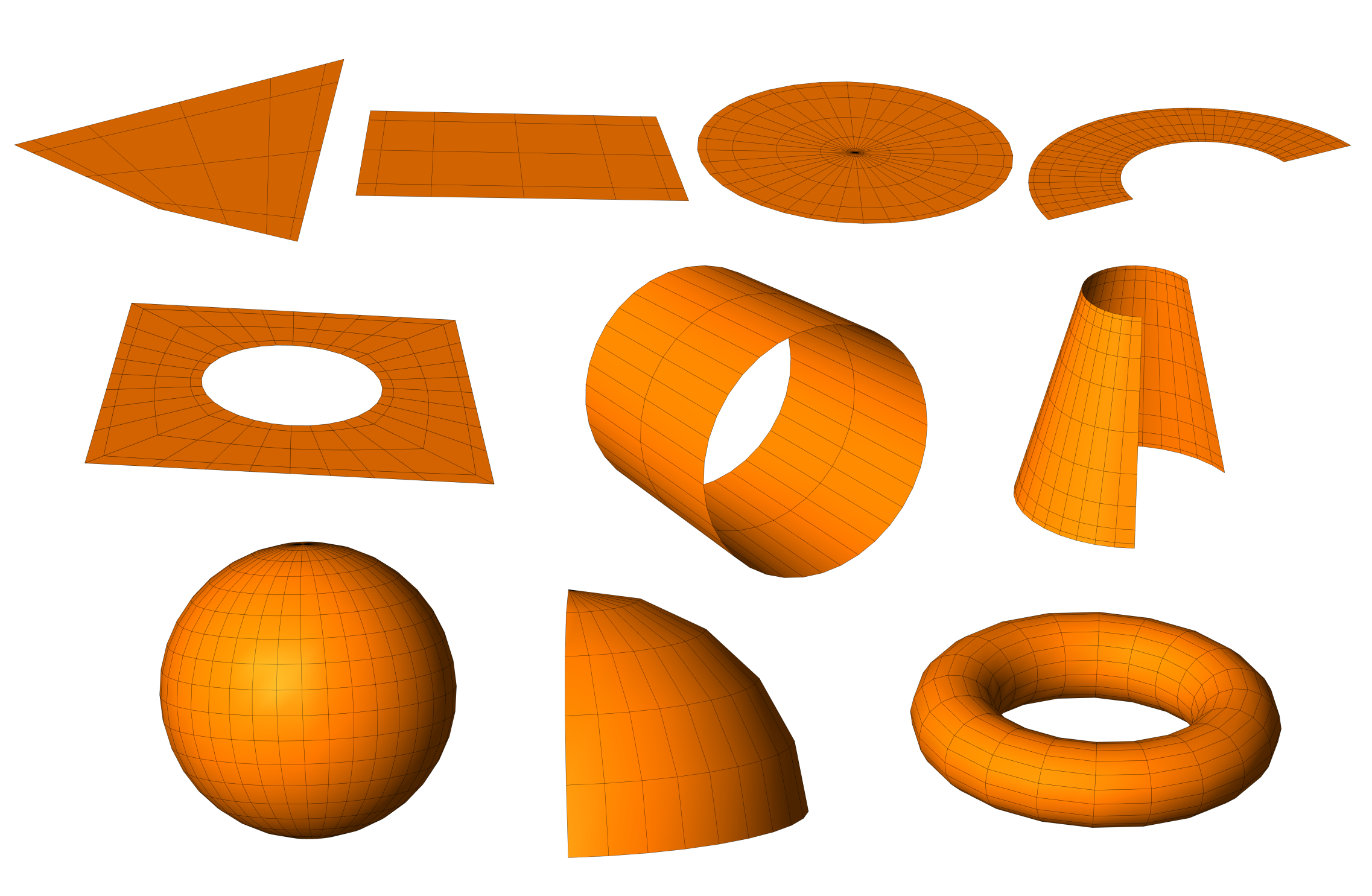}
\caption{{\textsc{gmsh}} view of the basic shapes available in \our.}
\label{fig:setshapes}
\end{figure}

As required by BEM, each volume is represented by its enclosing surfaces which in most cases can be split into basic shapes. The following basic shapes are currently available: quadrilateral through the coordinates of its corners, rectangular plate, disc, circular annulus (ring or crown), cylinder and cone, as well as sphere and torus. When necessary, only sectors of these shapes can be build such as for example an eighth of a sphere or an angular sector of a truncated cone as shown in figure~\ref{fig:setshapes}. In addition, a dedicated function to easily generate a square plate with circular hole has been implemented and more complicated shapes can be imported as meshes of triangular or quadrangular polygons both from \gmsh~\citep{Geuzaine:2009} and \catia (Dassault Syst\`emes) software. Several individual shapes labelled with an identical tag can be grouped into containers or objects in order to ease the application and/or modification of applied voltages during the simulation.

In order to build more complicated set-ups out of the basic shapes, some transformations (translation, rotation and reflection) can be applied to place the different shapes in their environment. In addition, up to three symmetry planes can also be included. The orientation of the normal to the surface can be chosen with the usual convention that it should point outward of a given volume. Both {\textsc{gmsh}} and {\textsc{root}}~\citep{Root:1997} may be used to display the geometry. 

\begin{figure}[htb!]
\begin{subfigure}{.225\textwidth}
  \centering
  \includegraphics[width=.95\linewidth,trim={0 0 5mm 5mm},clip]{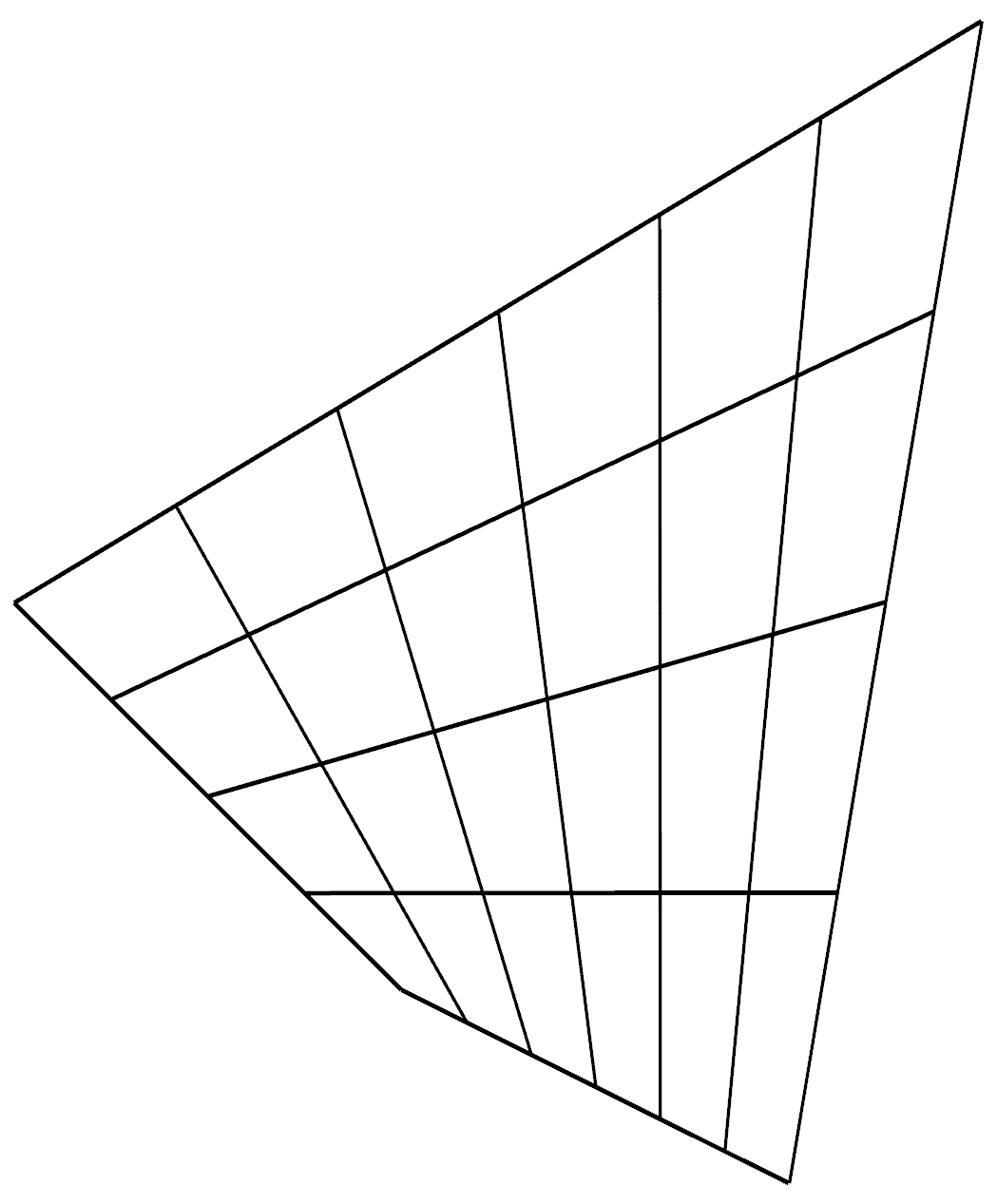}  
  \caption{Regular grid mesh in both directions. \\ {\ } \\ {\ }}
  \label{fig:rectangle_plate_RURU}
\end{subfigure}
\hfill
\begin{subfigure}{.225\textwidth}
  \centering
  \includegraphics[width=.95\linewidth,trim={0 0 5mm 5mm},clip]{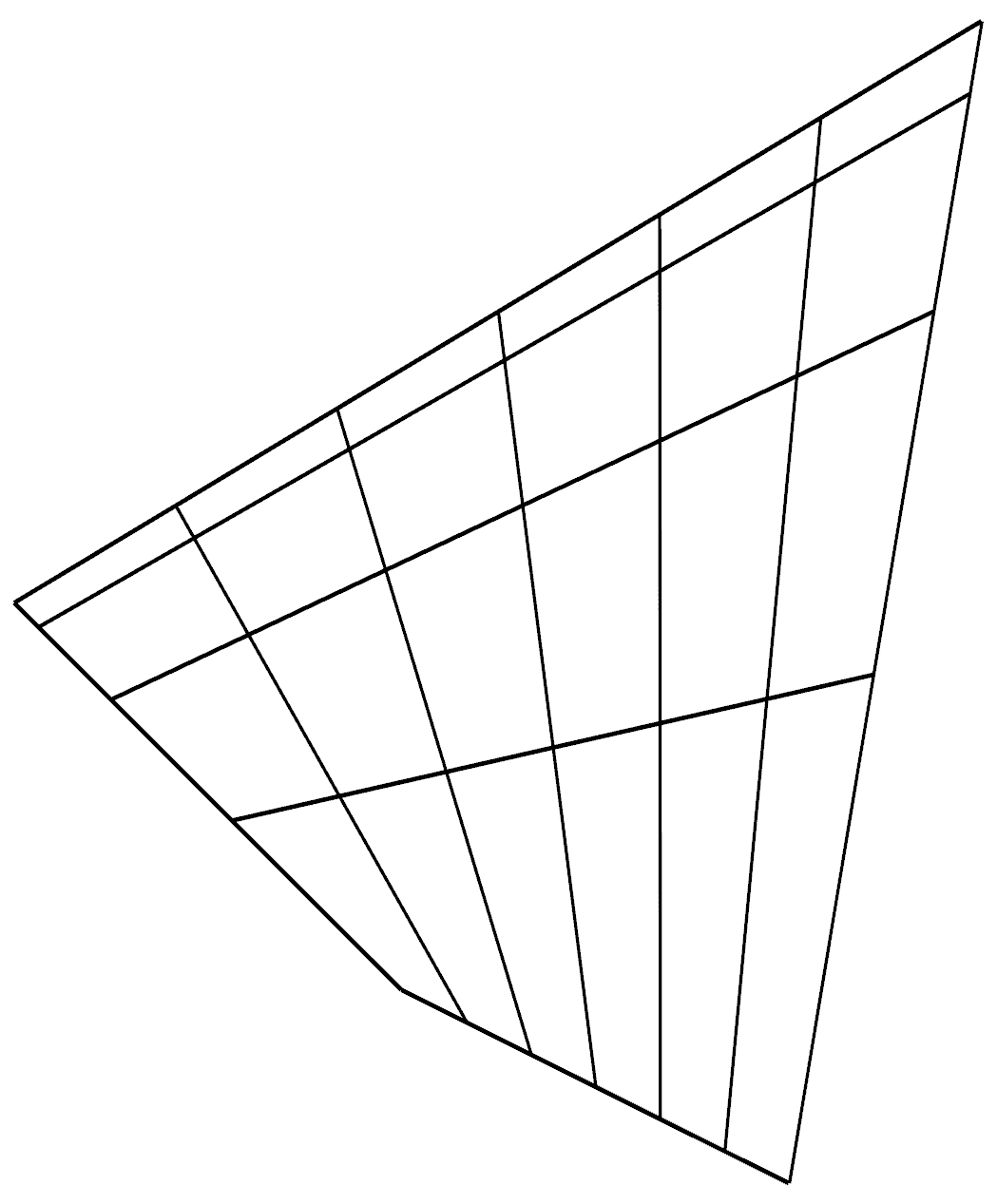}  
  \caption{Regular mesh in one direction and asymmetric geometric progression in other direction.}
  \label{fig:rectangle_plate_RUPU}
\end{subfigure}
\hfill
\begin{subfigure}{.225\textwidth}
  \centering
  \includegraphics[width=.95\linewidth,trim={0 0 5mm 5mm},clip]{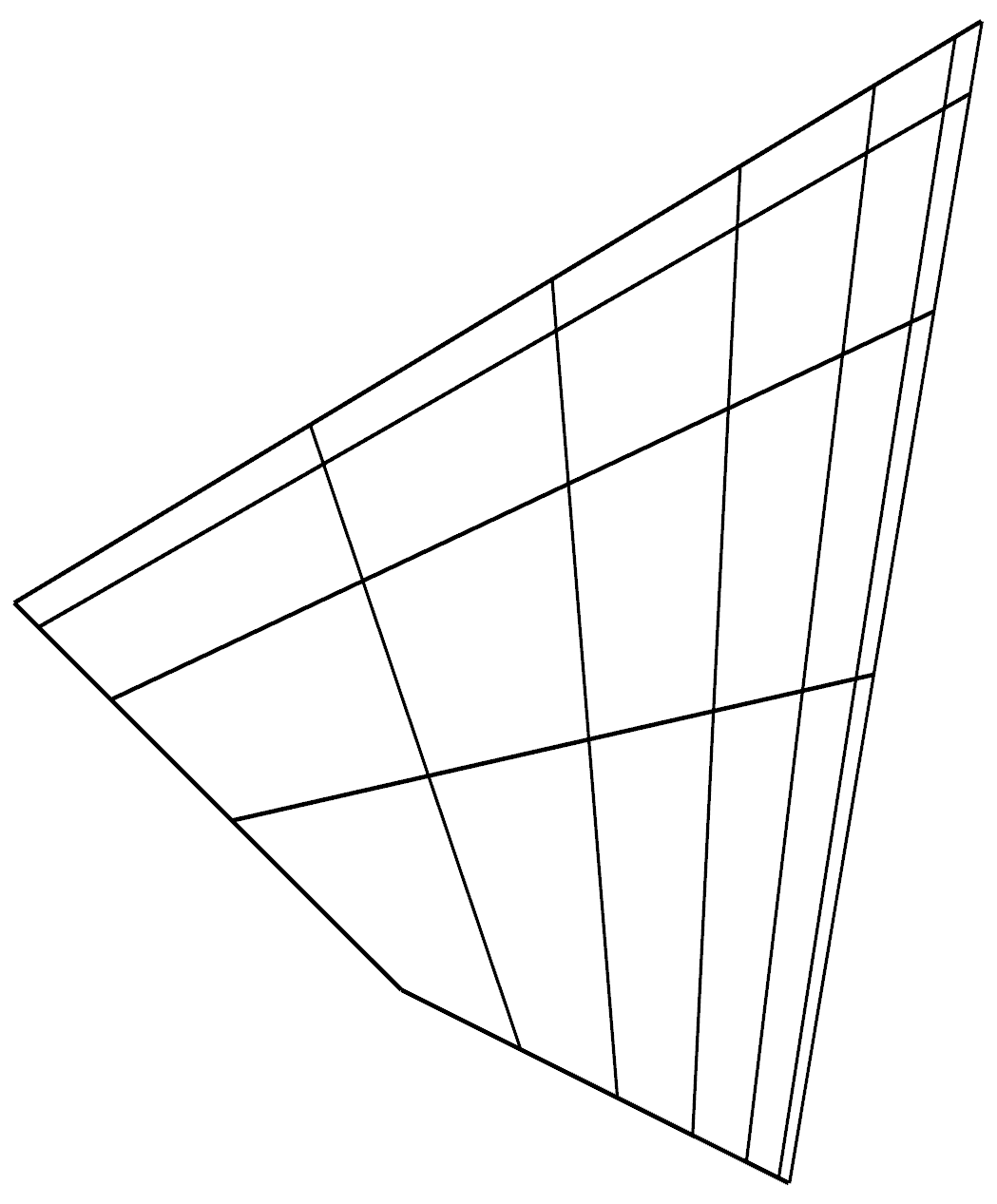}  
  \caption{Asymmetric geometric progression mesh in both directions. \\ {\ }}
  \label{fig:rectangle_plate_PUPU}
\end{subfigure}
\hfill
\begin{subfigure}{.225\textwidth}
  \centering
  \includegraphics[width=.95\linewidth,trim={0 0 5mm 5mm},clip]{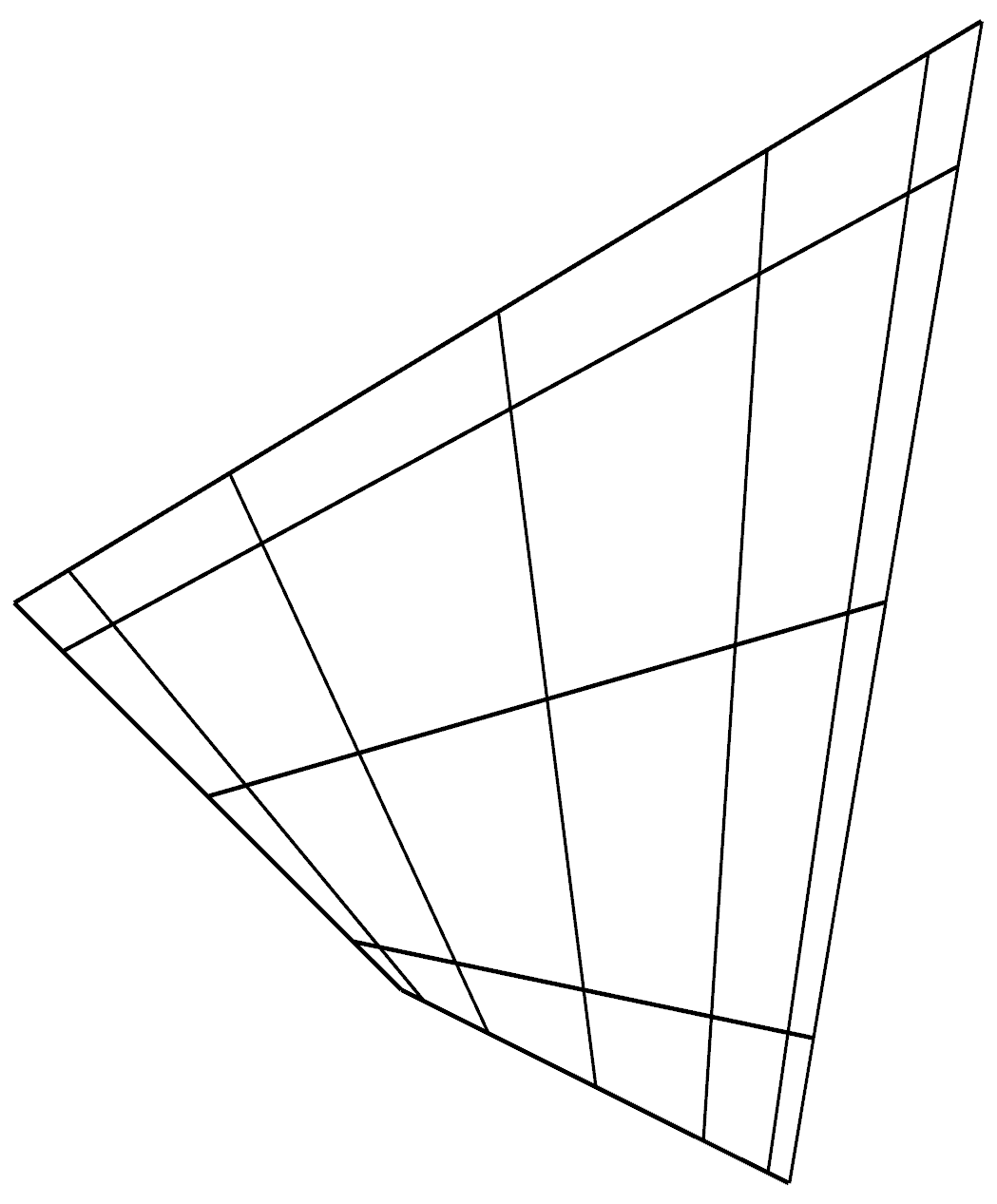}
  \caption{Symmetric geometric progression mesh in both directions. \\ {\ }}
  \label{fig:rectangle_plate_PSPS}
\end{subfigure}
\caption{Examples showing how the size and shape of the cells can be controlled in the mesh of a quadrilateral. The black lines demarcate the different cells.}
\label{fig:MeshOptions}
\end{figure}

Once the geometric entities have been defined, \our converts all surfaces into a mesh in which the shapes are approximated by a set of triangular and quadrilateral cells. This mesh is required by the BEM method to solve the electrostatic problem. The number of cells as well as their size and shape may strongly influence the precision of the computed field that can be reached by the solver. Adaptive mesh refinement is not (yet) supported in \our: the mesh properties are fixed when the geometry is declared. Several options are however available to precisely control the mesh characteristics. As shown in figure~\ref{fig:MeshOptions}, the mesh can be regular (or transfinite) with cells having the same size along  given sides of the geometrical entities or can follow some geometric progression. This last approach is useful to produce smaller cells where the surface charges are large or vary rapidly as is the case near sharp edges due to the point effect.

%
%
\subsection{Potential and field outputs and computation performances}
\label{sec:Perfs}

The knowledge of the mesh geometry along with the applied potentials and dielectric permittivities allows to fill and solve the system of \eqref{eq:matrixBEM} and to obtain the potential and field components everywhere in space. Once this has been achieved, a complete set of plotting/mapping functions allows to display and/or export the potential and field components in 1D, 2D or 3D.

Computing the potential and vector field on a given grid can be quite time consuming especially when the number of grid points is large. Table~\ref{table:StaticFieldTiming} gives examples of the computation times needed to evaluate the electric field components on several numbers of grid points using one or more GPUs for 8192 cells. For this, the geometry of the system was simply composed of two parallel plate electrodes separated by a vacuum gap. The maximum number of grid points is however limited by the amount of memory available on the GPU card. On a Tesla V100 with 32~GiB of DRAM, the limit is 512$^2\times$1024, i.e.~approximately 2.7$\times$10$^8$ points.

\begin{table}[!h]
    \centering
    \caption{\label{table:StaticFieldTiming} Computation times for the evaluation of the electric field components on several numbers of grid points using one or more GPUs for 8192 surface mesh elements.}
	\renewcommand{\arraystretch}{1.1}
	{\begin{tabular}{crr}
    	\toprule
        \# grid points & 1 GPU & 4 GPUs\\ \midrule
    	256$^3$ & 37~s &  9~s\\
    	512$^3$ & 212~s & 73~s\\
    	512$^2\times$1024 & 578~s & 146~s\\ \bottomrule
	\end{tabular}}
\end{table}

Adding more GPUs linearly reduces the computation times from around 10~min with one GPU to 2.4~min with 4 GPUs in the case of the maximum number of points.

%
%
\section{Charged particle transport}
\label{sec:MicroSim}

One of the main goals of \our is to study the dynamics of charged particles within gas media in presence of an electric field as is the case for instance in avalanche detectors used in particle physics. For this purpose, in addition to the geometry implementation which has been thoroughly described in the previous section, one needs to generate an initial set or cloud of particles with given characteristics. Then these particles are propagated by integrating their equations of motion with the help of a stepper algorithm fed by some electric field components in order to compute the forces acting on each particle. In addition, during their path through the gas medium, the particles will encounter some interactions. Starting with the generation of particles, all these ingredients shall be detailed in subsequent sections.
%
%
\subsection{Generation of initial particles}
\label{ssec:PartGen}

Initial particles, both electrons and ions, can be generated in the simulation according to different scenarios. The user can choose whether to: {\it i)} read an external file where the number of particles and their properties are defined such as the position, velocity vector, charge, mass and kinetic energy; {\it ii)} use one of the provided random generators; {\it iii)} distribute the particles along the path of primary particle trajectories. In these two latter cases, which we shall detail in the following paragraphs, particles are generated by pairs of an electron and an ion of the gas species.

%
%
\subsubsection{Particle random generators}
\label{sssec:PartRandom}

\begin{figure}[htb!]
     \begin{subfigure}[b]{0.51\textwidth}
     \caption{}
         \centering
          \includegraphics[width=\linewidth,trim={0 0 5mm 5mm},clip]{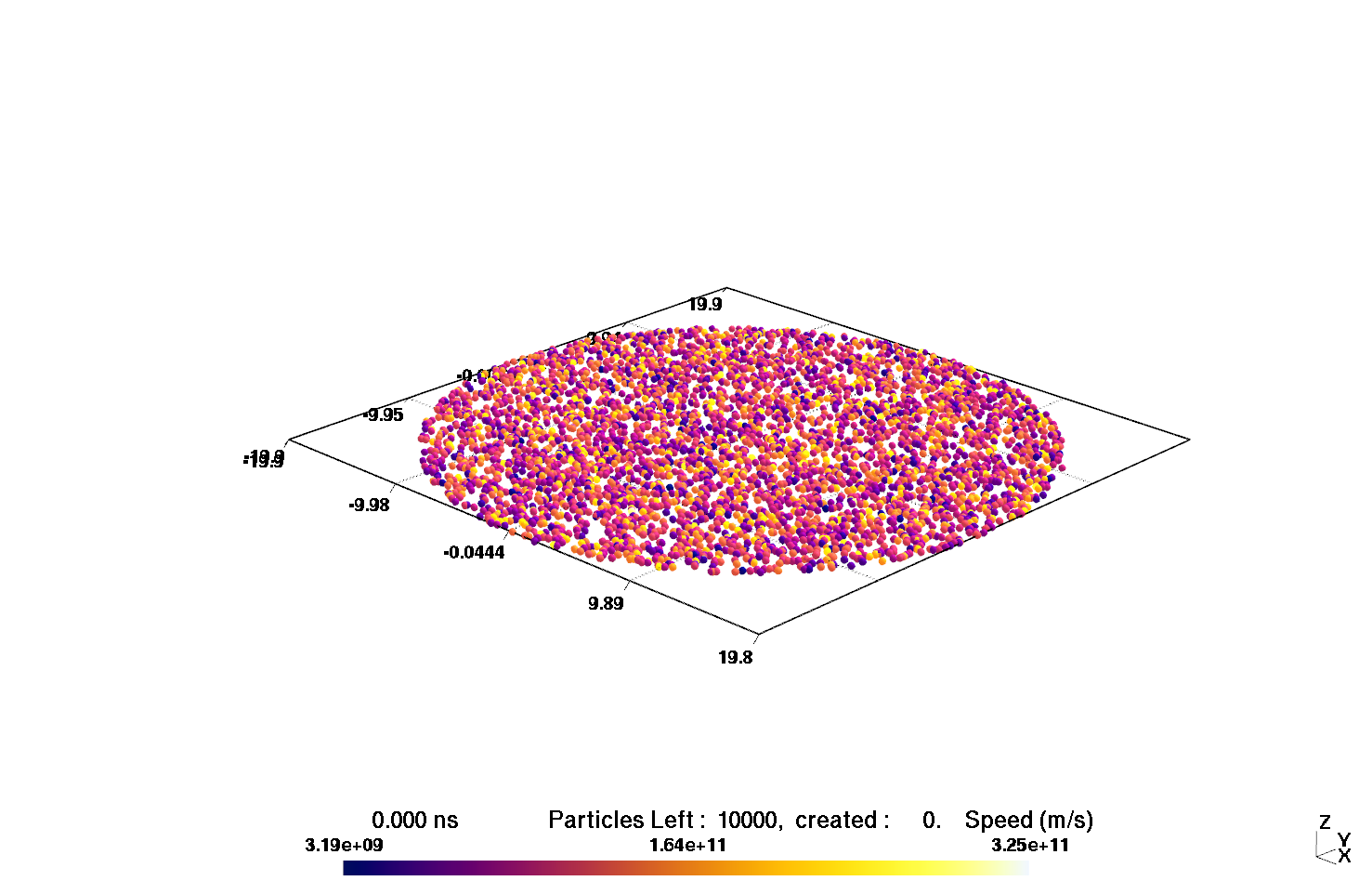}
         
     \end{subfigure}
     \hfill
     \begin{subfigure}[b]{0.51\textwidth}
         \caption{}
         \centering
         \includegraphics[width=\linewidth,trim={0 0 5mm 5mm},clip]{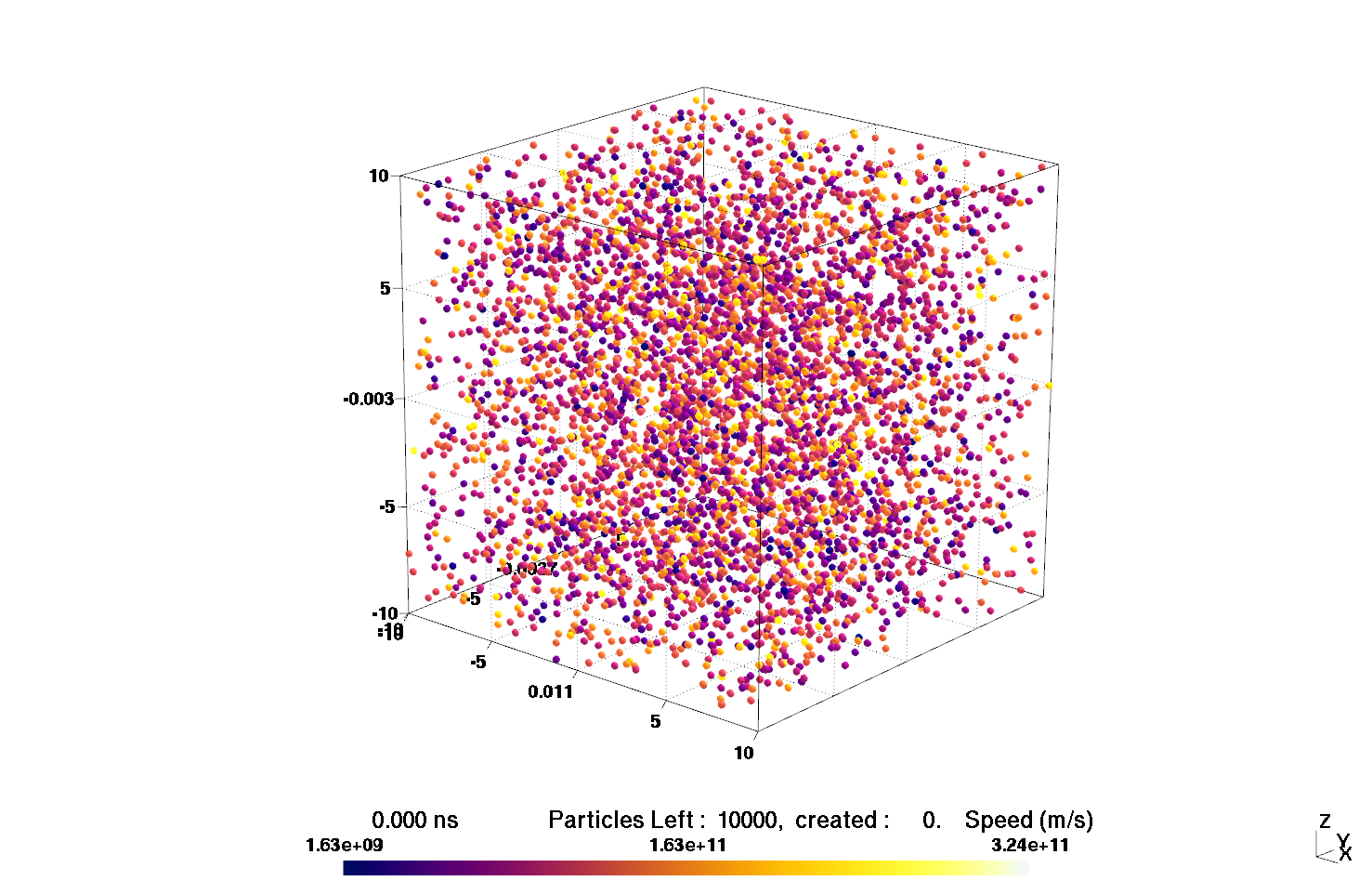}
     \end{subfigure} \\ \\
     \begin{subfigure}[b]{0.51\textwidth}
         \caption{}
         \centering
         \includegraphics[width=\linewidth,trim={0 0 5mm 5mm},clip]{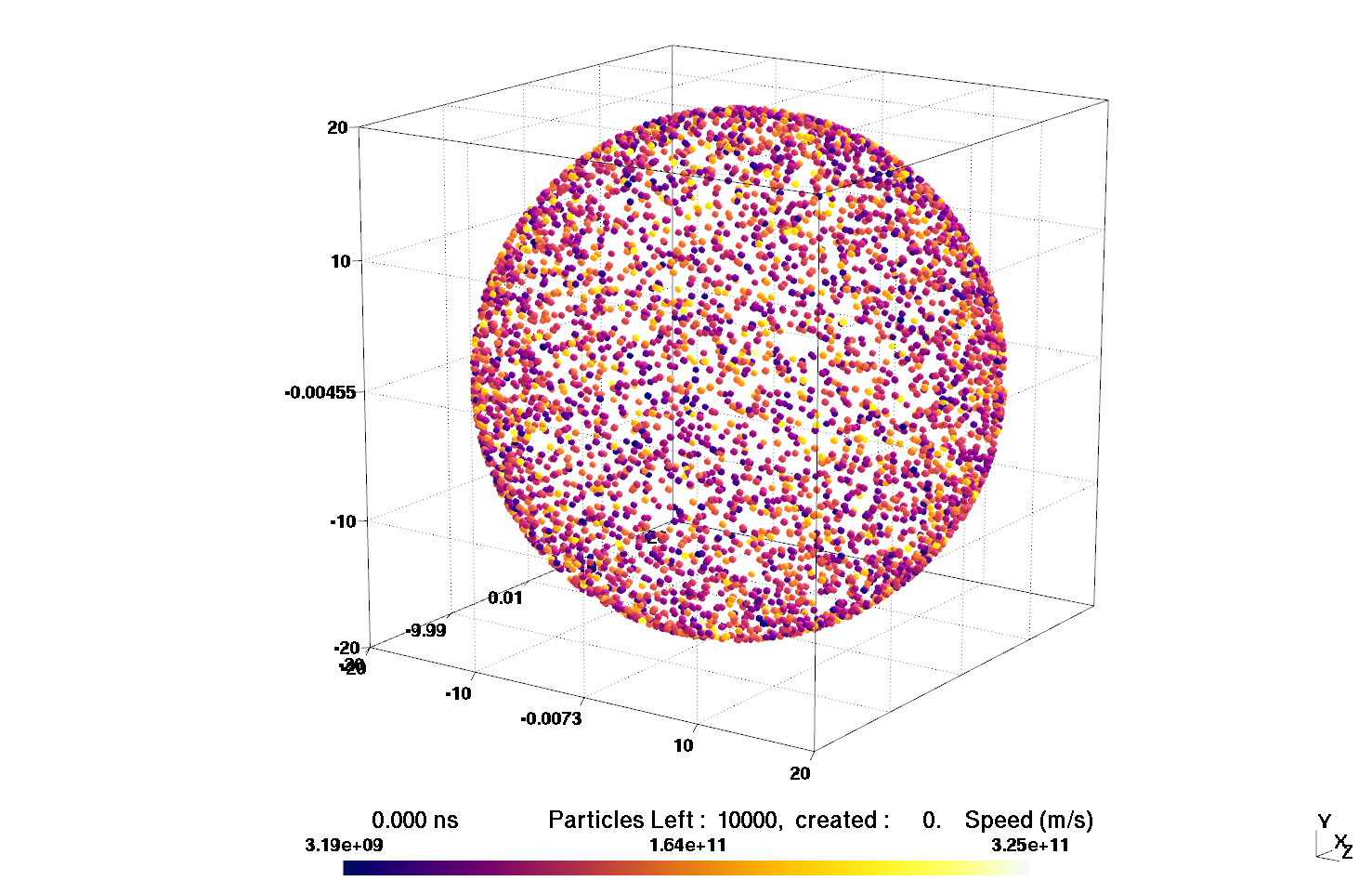}   
     \end{subfigure}
     \hfill
     \begin{subfigure}[b]{0.51\textwidth}
         \caption{}
         \centering
         \includegraphics[width=\linewidth,trim={0 0 5mm 5mm},clip]{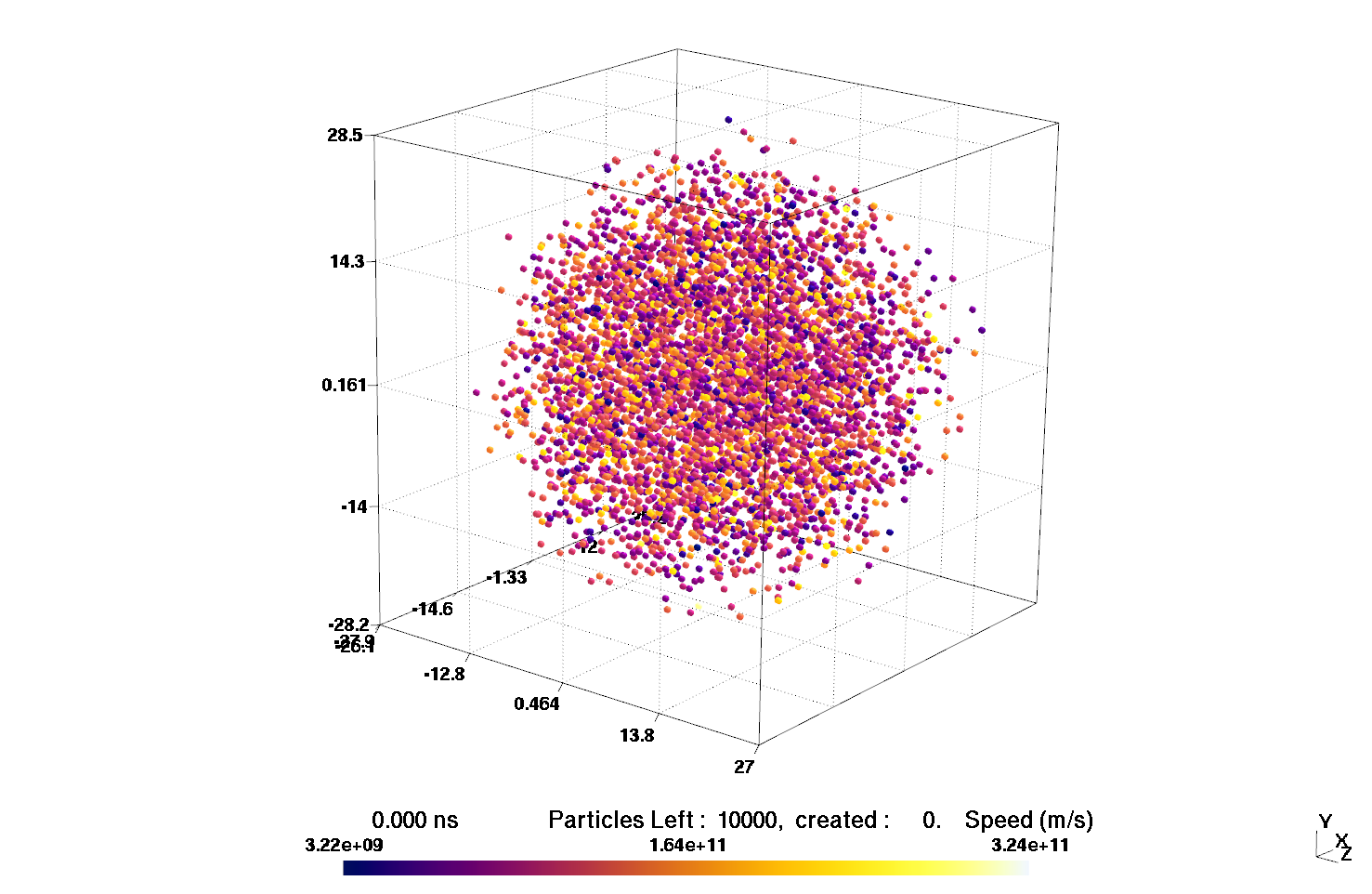}
     \end{subfigure}%
\caption{Examples of particle positions randomly distributed over: (a) a disk, (b) a cube, (c) a sphere and (d) a Gaussian ball. The colour scale shows the particle velocities in m/s.}
\label{fig:GenPart}
\end{figure}

The initial position and velocity vectors of the particles can be randomly generated according to some pre-defined probability distribution functions (PDFs).
The positions can have a uniform or a Gaussian-like distribution over surfaces such as squares, disks, cylinders, spheres, etc. or within volumes like for instance cubes, tubes, rods, balls and so on.
Different PDFs are available to generate the velocity vectors. For example, these vectors can be isotropically distributed  with a constant modulus based on a given kinetic energy or distributed following a Maxwell-Boltzmann distribution.
As an illustration, figure~\ref{fig:GenPart} shows the positions of 10$^4$ particles generated according to different PDFs. The velocity modulus follows a Maxwell-Boltzmann distribution and is colour coded in the figure.

%
%

\subsubsection{Primary particle trajectories}
\label{sssec:PartPrimary}

\begin{figure}[htb!]
    \centering
    \includegraphics[width=.75\linewidth]{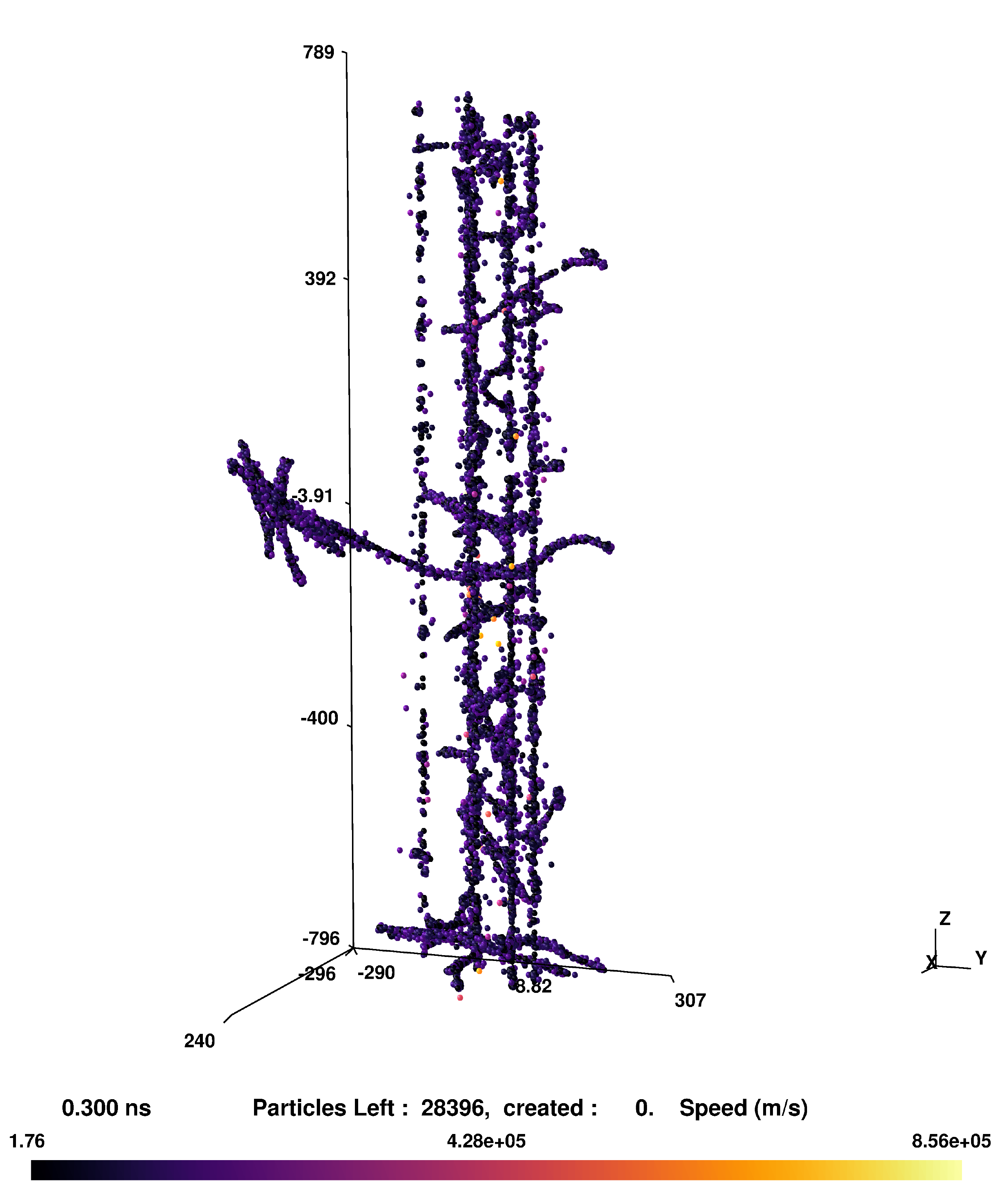}
    \caption{Positions of the 28396 particles produced after 300~ps by 5 primary carbon ions with 100~M$e$V/A in a 1.6~mm thick Ar/CO$_2$ (70/30)\% gas volume at 1~bar depositing 324~k$e$V. The colour scale represents the velocities of the particles in m/s.}
    \label{fig:tracks}
\end{figure}

When specifying that a primary track is at the source of the particles generation, the type (electron, proton, alpha or carbon), the number, the normalised momentum, the position and the kinetic energy of the primary track particles are set by the user. The energy deposited in the gas volume is calculated using the Bethe formula. The energy straggling is then taken into account using either the Landau-Vavilov or the Gauss formalism whether the gas volume is considered thin or thick~\citep{leo}. The number of particles created is calculated using the average energy for pair creation in the gas and their positions are randomly distributed following a uniform distribution along the primary tracks. The initial kinetic energy $E_K$ of the electrons of the pair is randomly distributed following the probability density function (PDF) given by~\citep{Cucinotta:1996}:

\begin{equation}
   \mathrm{PDF}(E_K) = C_{bb} \rho \frac{Z}{A} \frac{z^2}{\beta^2}\frac{1}{E_K^2}\left(1 - \beta^2 \frac{E_K}{Q_{\mathrm{max}}} + \frac{\pi\beta z^2}{137} \sqrt{\frac{E_K}{Q_{\mathrm{max}}}} \left(1 - \frac{E_K}{Q_{\mathrm{max}}}\right)\right).
   \label{eq:deltaRays}
\end{equation}

This function allows the production of $\delta$-rays where $Q_{\mathrm{max}}$ is the maximum energy transfer, $\beta$ is the reduced velocity and $z$ the effective charge of the primary particle, $Z$ and $A$ are the atomic charge and mass of the interaction medium, $\rho$, is the density of the medium and $C_{bb} = 0.1535$~M$e$V~cm$^2$/g, is the Bethe-Bloch coefficient.

The electron diffusion angles are such that the azimuthal angle $\phi$ is randomly distributed in $2\pi$~rad while the scattering angle $\theta$ follows:

\begin{equation}
   \cos^2\theta = \frac{E_K}{Q_{\mathrm{max}}}.
   \label{eq:theta}
\end{equation}

Even though the production of low energy electrons in a gaseous medium is sufficiently accurate to be used in the following processes, this method cannot replace a full Monte Carlo simulation for particle interactions such as Geant4~\citep{AGOSTINELLI2003250}.

An example of the secondary particles produced in a 1.6~mm thick Ar/CO$_2$ (70/30)\% gas volume at 1~bar by 5 primary carbon ions with energies of 100~M$e$V/A is shown in figure~\ref{fig:tracks}. The particles are shown after 300~ps evolution in the simulation.

\vspace{0.5cm}
Now that the initial set of particles has been generated, \our is able to proceed to the transport phase of the simulation as presented in the following section.

%
%

\subsection{Particle transport}
\label{ssec:PartTrans}

The charged particles transport requires the knowledge of the total force acting on each particle which is calculated from the electric field at particle location. In \our this electric field is the sum of four contributions (some of them being optional upon user's choice): {\it{i)}} the overall field due to both the applied potential and the presence of dielectric media in the set-up; {\it{ii)}} the field due to the influence of the particles on the charge density distributions at the $ed$- and $dd$-interfaces; {\it{iii)}} the charging-up effect when particles stick to the dielectric media surfaces; {\it{iv)}} the Coulombian field due to the charged particles themselves, this is the so-called N-body problem. Each of these contributions is described in more details in the following sections. The sum of these contributions is then used to integrate the equations of motion time step after time step.

%
%

\subsubsection{Field from the set-up}
\label{sssec:DynamicField}

Two options are available in \our to compute the electric field from the electrode voltages and the dielectric properties of the considered media: a \emph{static} and a \emph{dynamic} field option. When selecting the \emph{static} field option, the field components are computed only once at some predefined spatial coordinates located on a uniform 3D grid map before starting the transport of the particles. The grid size and refinement is left to the user's appreciation. During the transport, the field components at each particle location are then interpolated in this 3D map. Even though the \emph{static} field option implies to neglect the influence of the particles and the charging-up effect on the field distribution (see sections~\ref{sssec:Influence} and \ref{sssec:Charge}), it is much faster than using the \emph{dynamic} field option.

This later however, requires to solve the system~\eqref{eq:matrixBEM} and then to compute the field with~\eqref{eq:BEM3} at each $\Delta t$ time step. Nevertheless, even if slower, the \emph{dynamic} field option has the advantage to allow the use of time dependent applied voltages which can be updated at each time step. With this option, the potential of an electrode or of a set of electrodes can be varied according to some time dependent functions. Currently available functions include constant potential, linear or exponential variations as well as Fourier series expansions with a specified list of harmonics. The cell potentials $V_i$ of the electrode(s) concerned by such a variation are modified and new values of the $\sigma_i, \, i \in [1, \mathcal{N}]$ are computed. This is particularly interesting when dealing with radio-frequency potentials or the depolarisation of the system.

%
%

\subsubsection{Influence of the particles}
\label{sssec:Influence}

In some systems, and this is especially the case for gaseous detectors, it is important to take into account the influence of a large number of charged particles $\mathcal{N}_p$ within the simulated configuration, being part of the so-called space charge effects. In fact, these particles generate the averaged potential $\bar{V}$ and field $\bar{\mathbf{E}}$ over the surface of cell $\mathcal{C}_i$ following:
\begin{eqnarray}
  \bar{V}_{\textrm{particles} \rightarrow \mathcal{C}_i} & = & \frac{1}{4\pi\varepsilon_0} \sum_{p=1}^{\mathcal{N}_p} \frac{q_p}{A_i} \, \iint \limits_{\mathcal{C}_i} \frac{1}{\| \mathbf{r} - \mathbf{r}_p \|} \mathrm{d}^2\mathbf{r}, \label{eq:potpart}\\
  \bar{\mathbf{E}}_{\textrm{particles} \rightarrow \mathcal{C}_i}& = & \frac{1}{4\pi\varepsilon_0} \sum_{p=1}^{\mathcal{N}_p} \frac{q_p}{A_i} \, \iint \limits_{\mathcal{C}_i}  \frac{\mathbf{r} - \mathbf{r}_p}{\| \mathbf{r} - \mathbf{r}_p \|^3} \mathrm{d}^2\mathbf{r}, \label{eq:fldpart}
\end{eqnarray}
 with $A_i$ the area of cell $\mathcal{C}_i$ and $q_p$ the charge of particle $p$ located at position $\mathbf{r}_p$. For cells with small dimensions, the field due to the  $\mathcal{N}_p$ particles could be evaluated at the cell centre, but for larger cells the average over the whole cell area is a more reasonable approximation. The presence of these particles will modify the charge distributions on the electrodes and dielectric-to-dielectric interfaces. Left hand sides of \eqref{eq:BEMpot}, \eqref{eq:BEMfld} and \eqref{eq:matrixBEM} should therefore be modified accordingly to account for this effect before solving:
\begin{eqnarray}
   V_i \, - \, \bar{V}_{\textrm{particles} \rightarrow \mathcal{C}_i} & = & \sum_{j=1}^{\mathcal{N}} \, Q_{ij} \, \sigma_j \qquad i \in [1, \mathcal{N}_{ed}],\label{eq:BEMpotpart} \\
\sigma_{\mathrm{F},i} \, - \, (\varepsilon_2 \, - \, \varepsilon_1) \, \bar{\mathbf{E}}_{\textrm{particles} \rightarrow \mathcal{C}_i} \cdot \mathbf{\hat{n}}_i & = & \sum_{j=1}^{\mathcal{N}} \, W_{ij} \, \sigma_j \qquad i \in [\mathcal{N}_{ed}+1, \mathcal{N}],\label{eq:BEMfldpart}
\end{eqnarray}
with $\mathbf{\hat{n}}_i$ the normal vector to cell $\mathcal{C}_i$.
Once the $\mathcal{N}$ corresponding charge densities $\sigma_j$ have been solved for, the potential and field at any location $\mathbf{r}$, and therefore at the $\mathcal{N}_p$ particle positions, are obtained with~\eqref{eq:BEM2} and \eqref{eq:BEM3}.

%
%

\subsubsection{Charging-up}
\label{sssec:Charge}

Charging-up is a phenomenon in which charged particles, both electrons and ions, are collected by dielectric material surfaces. Because of the insulating properties of the dielectric materials, these charges cannot escape to nearby electrodes and thus modify the surrounding surface charge densities resulting in a distortion of the overall electric field. A similar phenomenon also happens in the case where electrodes have a non negligible surface resistivity, the collected charges contribute to the surface charge density until they are slowly evacuated with some specific decay time. 

To take the charging-up effect into account, let us assume that $\mathcal{N}_{p,i}$ charged particles with charge $q_p$ ($p \in [1, \mathcal{N}_{p,i}]$) are collected by the $dd$-cell $\mathcal{C}_i$ during the time step $\Delta t$ and that these charges are uniformly distributed over the whole cell surface of area $A_i$. The free surface charge density $\sigma_{F,i}$ in the left hand side of \eqref{eq:matrixBEM} is therefore modified in the following way:
\begin{equation}
     \sigma_{F,i}(t) = \sigma_{F,i}(t - \Delta t) + \frac{1}{A_i} \sum_{p=1}^{\mathcal{N}_{p,i}} q_p.
     \label{eq:charge-up}
\end{equation}
Such a modification is applied to all $dd$-cells collecting charges and the system~\eqref{eq:matrixBEM} is solved again for the new $\sigma_i$ on both the $ed$ and $dd$ interfaces.

%
%

\subsubsection{The N-body problem}
\label{sssec:NBody}

Each of the $\mathcal{N}_p$ charged particles present in the set-up at a given time generates a Coulombian attraction or repulsion on each of the other particles.
This is the so-called N-body problem where the field on a particle $p^\prime$ located at the position $\mathbf{r}_{p^\prime}$ due to all the other particles is given by:
\begin{equation}
  \mathbf{E}_{\textrm{N-body}}(\mathbf{r}_{p^\prime}) \, = \, \frac{1}{4\pi\varepsilon_0} \sum_{\substack{p=1\\ p\neq p^\prime}}^{\mathcal{N}_p} q_p \frac{\mathbf{r}_{p^\prime} - \mathbf{r}_p}{\| \mathbf{r}_{p^\prime} - \mathbf{r}_p \|^3} \qquad \forall \, p^\prime \, \in \, [1, \mathcal{N}_p].  \label{eq:nbody}
\end{equation}

This problem computation exhibits an $\mathcal{O}(\mathcal{N}_{p}^2)$ complexity which can lead to more than 10$^{13}$ floating point operations for $\mathcal{N}_{p} = 10^6$ particles every time step. 
For large numbers of particles, the brute force (BF) calculation which consists in a direct evaluation of~\eqref{eq:nbody}, leads to prohibitive computational times. This is why many approximation methods have been developed to decrease the complexity (and therefore the computational time) from $\mathcal{O}(\mathcal{N}_{p}^2)$ to $\mathcal{O}(\mathcal{N}_{p} \log\mathcal{N}_{p})$ or even $\mathcal{O}(\mathcal{N}_{p})$ such as for instance the Barnes-Hut method~\citep{BarnesHut:1986} or the Fast Multipole Method~\citep{FMM:1987}. Implementation of this last method in \our is currently under investigation, but up to now, in addition to the BF method, a nearest neighbour (NN) approximation method has been included for which the implementation on GPUs is easier while keeping an $\mathcal{O}(\mathcal{N}_{p})$ computing complexity. Described briefly, it consists in evaluating the exact Coulombian field following~\eqref{eq:nbody} for each of the closest particles to the particle of interest while the field from the particles further away is approximated by their centre of charges with a barycentric method.

\begin{figure}[htb!]
    \centering
    \includegraphics[width=.75\linewidth]{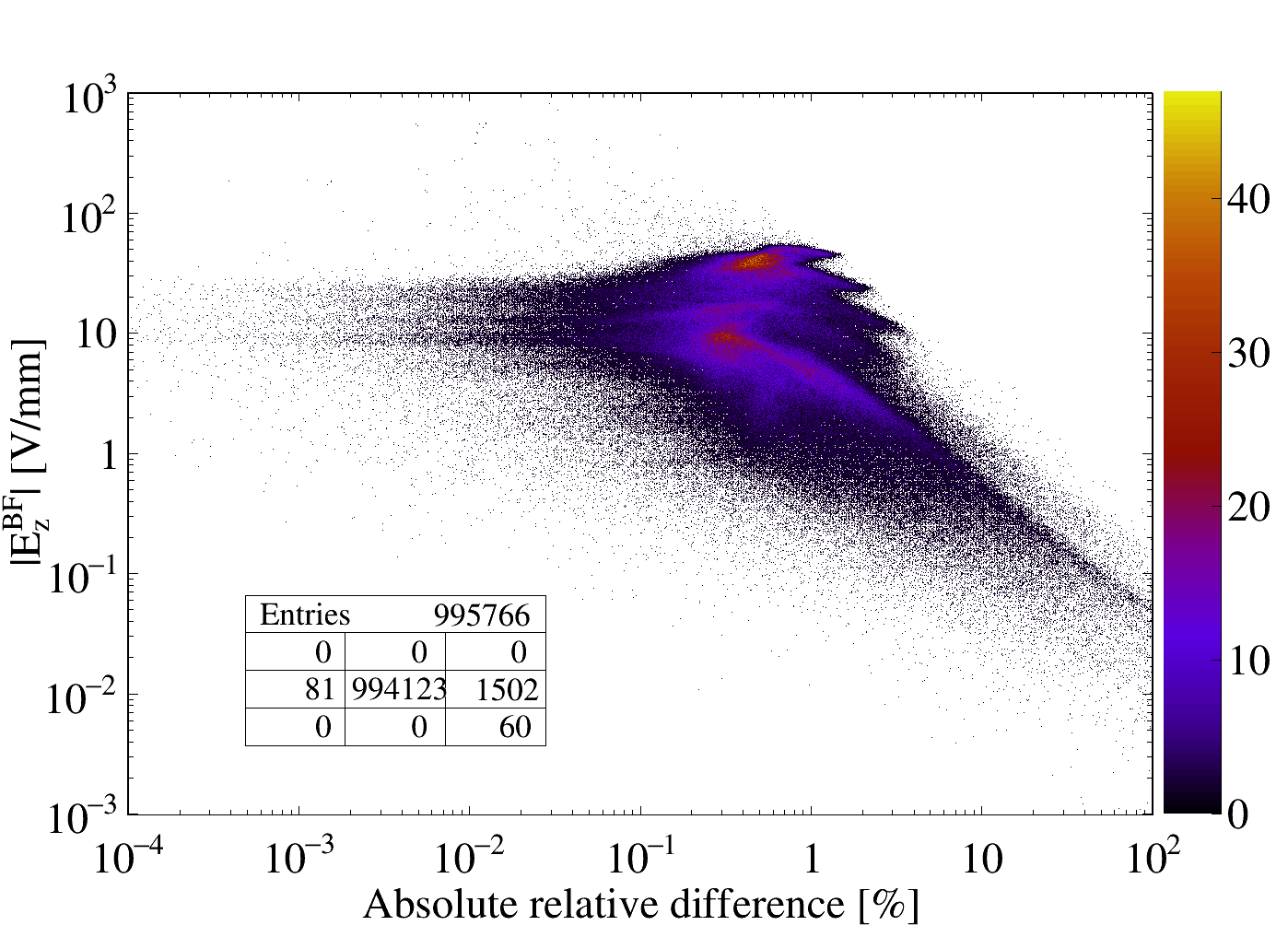}
    \caption{Absolute value of the field from the BF algorithm $|E_z^{\mathrm{BF}}|$ from brute force algorithm versus the absolute relative difference $|(E_z^{\mathrm{NN}} - E_z^{\mathrm{BF}}) / E_z^{\mathrm{BF}}|$ for the $z$ component of the field obtained with both methods at the location of $\sim 10^6$ charged particles (electrons + ions) at 2.5~ns of tracking, (i.e.~$10^5$ time steps). The bottom left box shows the numbers of particles lying inside and outside the axes ranges.}
    \label{fig:NBodyComp}
\end{figure}

Figure~\ref{fig:NBodyComp} presents, for a simple parallel plate avalanche detector, the absolute value of the field from the BF algorithm $|E_z^{\mathrm{BF}}|$ versus the absolute relative difference $|(E_z^{\mathrm{NN}} - E_z^{\mathrm{BF}}) / E_z^{\mathrm{BF}}|$ for the $z$ component of the field obtained with the BF and NN methods at the location of approximately $10^6$ electrons and ions at 2.5~ns of tracking, (i.e.~$10^5$ time steps). This corresponds to the worst situation for the NN algorithm where the ion and electron clouds are well separated after their drift in opposite directions. The simulation shows that the absolute relative differences are below 5\% for more than 96.8\% of the particles. It should be emphasized that, for the 3\% remaining particles, the field components are very small leading therefore to small contributions to the overall repulsion or attraction forces.
Table~\ref{table:NBodyTiming} gives some examples of computation times obtained with 1 or 4 GPU cards for both algorithms and several numbers of particles. The NN algorithm  clearly outperforms the BF one by more than 3 orders of magnitude when dealing with $10^7$ particles. This, by itself, justifies the use of the NN if very high precision is not needed in a given simulation.

\begin{table*}\centering
\caption{\label{table:NBodyTiming} Computation times for evaluating the N-body problem for one time step averaged over 100 time steps as a function of the number of particles for 1 and 4 GPUs.}
\ra{1.1}
\begin{tabular}{@{}ccrrcrr@{}}\toprule
\multirow{2}{*}{\# particles} &  & \multicolumn{2}{c}{BF} & \phantom{ab} & \multicolumn{2}{c}{NN} \\
\cmidrule{3-4} \cmidrule{6-7}
&& 1 GPU &  4 GPUs && 1 GPU & 4 GPUs\\ \midrule
		10$^5$ && 0.121~s & 0.05 s && 0.005~s & 0.004 s\\
		10$^6$ && 16~s    & 3.9~s  && 0.09~s & 0.03~s\\
		10$^7$ && $\sim$1600~s & 400~s && 1.1~s & 0.35~s\\
\bottomrule
\end{tabular}
\end{table*}

Once computed, the electric field from the N-body problem is then added to the other field contributions to obtained the corresponding total force acting on each particle. 
It has to be emphasized that on the contrary to the effects of influence and charging-up described in section~\ref{sssec:Influence} and \ref{sssec:Charge}, respectively, the N-body effect does not modify the charge densities on the geometry cells and therefore does not require to solve~\eqref{eq:matrixBEM}.

%
%

\subsubsection{Numerical integrators}
\label{sssec:Steppers}

Having the total electric field at each particle location and therefore the total force acting on each particle, a numerical integrator or stepper algorithm is used to integrate the equations of motion in order to update the velocity, the position and the energy of each particle every time step $\Delta t$.

Two stepper algorithms have been implemented in \our. The first one is the basic Euler algorithm which is very unstable and is known to have an error proportional to the time step. However, this algorithm has the great advantage to require only a single field evaluation during a time step $\Delta t$ and is therefore extremely fast. The second algorithm, called PEFRL~\cite{Omelyan:2002}, is a Leap frog-like fourth order symplectic solver which is much more stable and accurate and even time reversible. However, it requires 5 field evaluations during each $\Delta t$ which makes it very slow compare to the Euler stepper. It is therefore more suited for simulations performed with the "static field" option selected.

One should also notice that the \textit{memory} of the previous free flight steps is lost when the particle interacts with the gas medium (as depicted in the next section), thus counterbalancing the numerical integration error which accumulates time step after time step.

%
%
\subsection{Particle interactions with the gas medium}
\label{ssec:GasProp}

For detector simulations, a gas medium is added in the set-up. This gas is described by the list of its compounds or gas species, its temperature and its overall pressure. At the moment, only Ar~\citep{magboltz}, CO$_2$~\citep{CO2}, N$_2$~\citep{magboltz}, O$_2$~\citep{O2}, iC$_4$H$_{10}$~\citep{magboltz}, CF$_4$~\citep{CF4} and CH$_4$~\citep{CH4} can be included separately or as different mixtures. Depending on the gas type, the energy dependent electron-gas cross sections include elastic, inelastic, attachment and ionisation interactions. The program will then load the corresponding electron interaction cross sections scaled by the corresponding gas proportion and interpolate the missing data over an energy range from $E_{\mathrm{min}} = 1$~m$e$V to $E_{\mathrm{max}} = 1$~k$e$V. In order to accurately interpolate and include the data, a so-called normalised \textit{lethargy} parameter is used instead of the energy $E$. This parameter $\mathcal{L}(E)$ is defined as:

\begin{equation}
   \mathcal{L}(E) = \frac{\mathrm{ln}\left(\frac{E_{\mathrm{max}}}{E}\right)}{\mathrm{ln}\left(\frac{E_{\mathrm{max}}}{E_{\mathrm{min}}}\right)}.
   \label{eq:leth}
\end{equation}

\begin{figure}[htb!]
    \includegraphics[width=.9\linewidth,trim={0 0 0mm 0mm},clip]{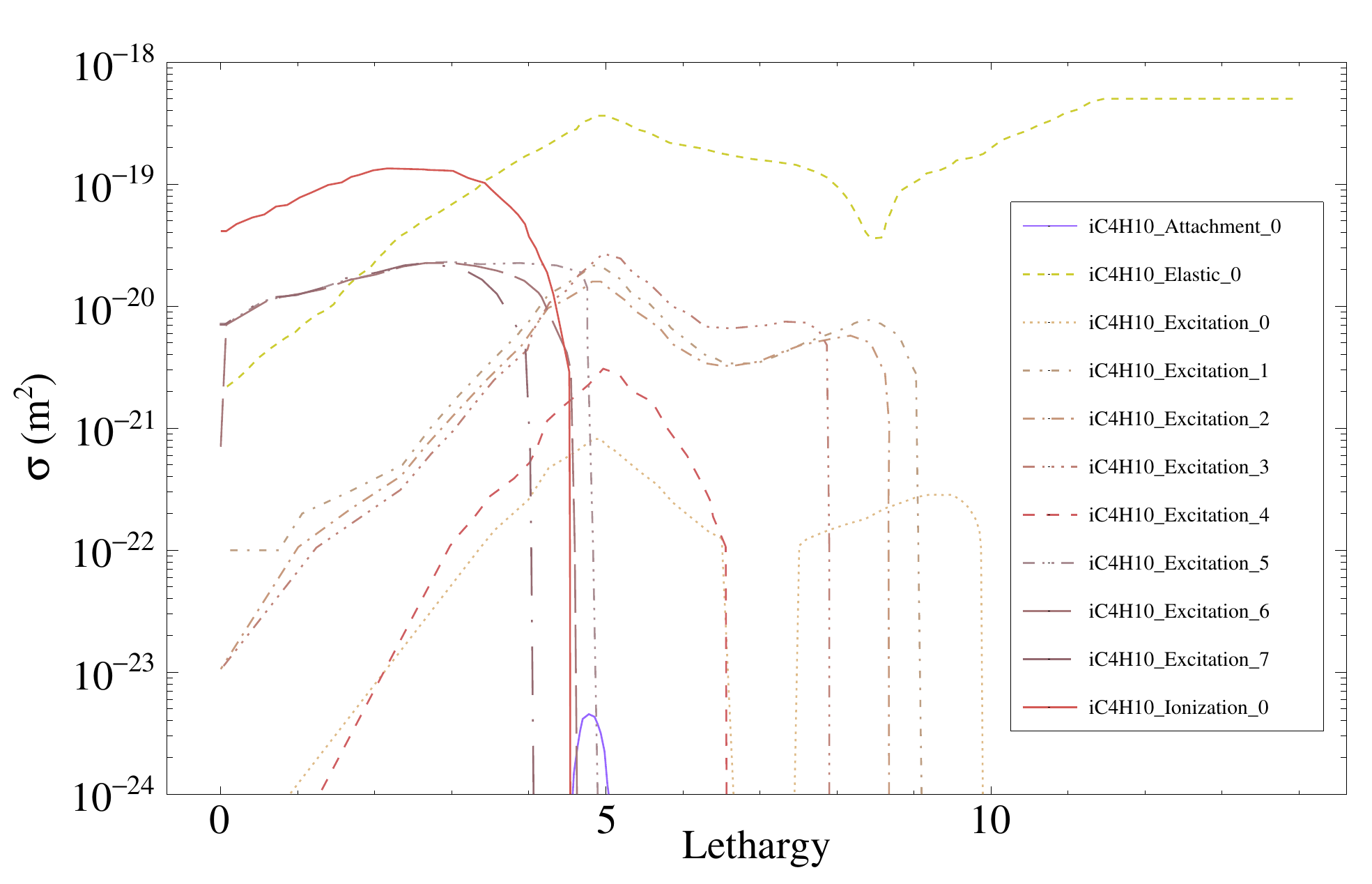}  
    \caption{Isobutane cross sections used in the program as a function of the \textit{lethargy} $\mathcal{L}(E)$. The cross sections shown here have been graphically extracted from~\citep{magboltz}.}
    \label{fig:XS}
\end{figure}

It allows for a linear variation over the whole energy range spanning 6 orders of magnitude.
Figure~\ref{fig:XS} shows as an example the cross sections used for the isobutane gas.

Concerning the ion interactions, the only available interaction cross sections currently included are the backscattering and isotropic cross sections of Ar$^+$ on Ar~\citep{Ar_Ar}. This means that the microscopic simulation of the ion interaction and transport is only accurate when choosing the argon gas and a temperature of 300~K.

%
%
\subsubsection{Gas interactions}
\label{sssec:Interactions}

Using the interaction cross sections of the different gas species included, the program computes every time step the probability for an electron to have an interaction with the gas, neglecting the velocities of its molecules and/or atoms compared to the electron velocity. For this, the time step must be chosen carefully in order to have much less than one interaction per time step. In the general case, the time step was chosen to be 25~fs giving for example an interaction probability for Ar/CO$_2$ (70/30)\% at 1~bar pressure lower than 0.1 with a collision frequency $\nu\simeq 10^{12}$s$^{-1}$. This method allows for an accurate evaluation of the collision rates without the need to use the null-collision technique~\citep{Skullerud_1968}. It is however more computationally expensive but is counter-balanced by the GPUs power.

The interaction type is then chosen randomly using a basic Markov-chain method. The interaction is treated as a two-body interaction. In the case of elastic interactions, anisotropic diffusion of the scattered electrons may arise depending on the gas species used. In this case, the diffusion angle $\theta$ of the electron is given by~\citep{Okhrimovskyy:2002}:
\begin{equation}
    \theta = \cos^{-1}\left( 1 - \frac{2R(1 - \xi)}{1 + \xi( 1 - 2R )}\right), \qquad \mathrm{where} \quad \xi(E) = f\left(\frac{\sigma_m(E)}{\sigma(E)}\right),
    \label{eq:anisotrop}
\end{equation}

where $E$ is the electron kinetic energy, $0\leq R < 1$, is a random number, $\sigma_m(E)$ and $\sigma(E)$, are the momentum transfer and total elastic cross sections, respectively.

%
%
\subsubsection{Penning effect}
\label{sssec:Penning}

The Penning effect describes the additional ionisation rates in a mixture of gas species due to excitation energy. When an electron undergoes an inelastic interaction with usually a noble gas atom, the excited state of this atom may react either through direct two-body collisions or photoemission with a different gas atom in the mixture, usually a molecular gas species having a lower ionisation potential, creating an electron/ion pair. In the program, these phenomena are taken into account using measured Penning transfer probabilities~\citep{PenningTransfer}. For example, in the case of an Ar/CO$_2$ (70/30)\% mixture, an electron/ion pair will be produced according to the Penning transfer probability (equal to 0.574 in this case) when an interaction occurred on an argon excited state with an energy higher than the ionisation potential of CO$_2$. As the ionisation potential for argon is higher than any excited state of CO$_2$, the opposite process cannot take place.

%
%
\subsubsection{Electron-ion recombination}
\label{sssec:Recomb}

Alongside to the attachment process where an electron is captured by a neutral atom of the gas to form a negative ion (anion), recombination is the process of a positive ion (cation) of the gas neutralized by an entity of opposite polarity. They do not participate anymore in the generation of the electrical signal and result in a loss of proportionality of the detector.

Out of the different recombination processes that can occur in a gas mixture (such as for instance one-step radiative recombination, two-step dielectronic recombination, three-body recombination, etc~\citep{Hahn_1997}), only the direct two-body electron-ion process has been included in the simulation which is an order of magnitude more likely to happen compared to three-body processes~\cite{IObodovskiy}. This phenomenon is taken into account microscopically for any electron and ion using the distances obtained during the N-body calculation process (see section~\ref{sssec:NBody}).

\begin{figure}[htb!]
    \centering
    \includegraphics[width=.5\linewidth]{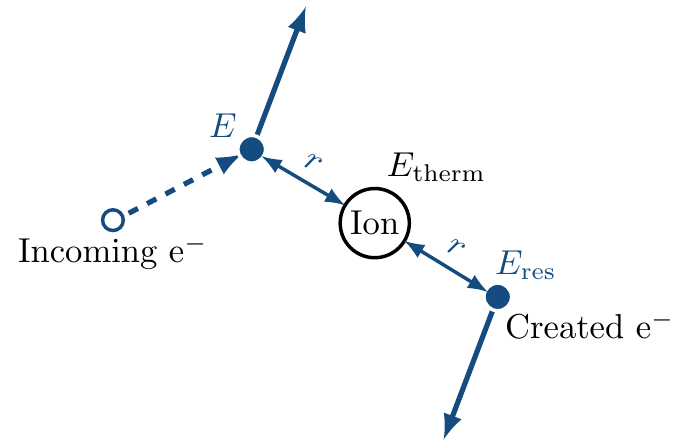}
	\caption{Schematic of the ionisation process in the simulation. The incoming electron ionises an atom of the gas and generates an electron/ion pair. The distance between the electrons and the ion is set according to~\eqref{eq:distIoni}. The ion is created with a thermal kinetic energy $E_{\mathrm{therm}}$ and the new electron has the residual energy $E_{\mathrm{res}}$ from the interaction.}
	\label{fig:coulomb}
\end{figure}

Assuming the classical view of the Bohr model on the ionisation process, when an ionisation interaction occurs from an electron with kinetic energy $E$, an electron/ion pair is produced from the ionisation potential energy $E_{\mathrm{ioni}}$. As shown in figure~\ref{fig:coulomb}, the ion is created with a kinetic energy corresponding to the thermal energy of the gas $E_{\mathrm{therm}}=\frac{3}{2}k T$ and the new electron acquires the residual kinetic energy such as $E_{\mathrm{res}}=E-E_{\mathrm{ioni}}-E_{\mathrm{therm}}$. The momentum in the center of mass is conserved so that both electrons have opposite velocities. The distance of both electrons to the ion is equal to $r$ given by~\cite{IObodovskiy}:

\begin{equation}
    r \, = \, \frac{1}{4\pi\varepsilon_0} \frac{e}{2E_{\mathrm{res}}}.
    \label{eq:distIoni}
\end{equation}

For a recombination between an electron and a cation to occur, the ionisation process is somehow reversed and the distance $d$ between both particles should be smaller than the distance of the Coulomb potential at the ionisation energy $E_{\mathrm{ioni}}$:
\begin{equation}
    d \leq \, \frac{1}{4\pi\varepsilon_0} \, \frac{e}{2E_{\mathrm{ioni}}}.
    \label{eq:distRecomb}
\end{equation}

In this case the particles recombine to form a neutral entity and are therefore removed from the simulation.

%
%

\subsection{Signal generation: Shockley-Ramo theorem}
\label{ssec:Ramo}

In order to generate the electric signals on the read-out electrodes, the formalism of the Shockley-Ramo theorem is used~\citep{ramo}. The current $i(t)$ induced on a read-out electrode from $\mathcal{N}_{p}$ moving particles of charge $q_p$ with instantaneous velocities $\mathbf{v}_p(t)$ is calculated as:

\begin{equation}
    i(t) = \sum_{p=1}^{\mathcal{N}_{p}} q_p \mathbf{v}_p(t)\cdot \mathbf{E}_w,
     \label{eq:ramo}
\end{equation}
where the so-called Ramo weighting field $\mathbf{E}_w$ is calculated, without any particles, under well-known conditions: the read-out electrode potential is set to 1~V, while all others electrodes are grounded. It must be calculated for each read-out electrode defined by the user. Two cases can be set-up to generate the weighting Ramo electric fields: either in the static option where the electric field for each read-out electrode is calculated beforehand and stored in field maps or in the dynamic option where it is calculated in real-time during the field calculation process.

\begin{figure}[htb!]
    \centering
    \includegraphics[width=.6\linewidth]{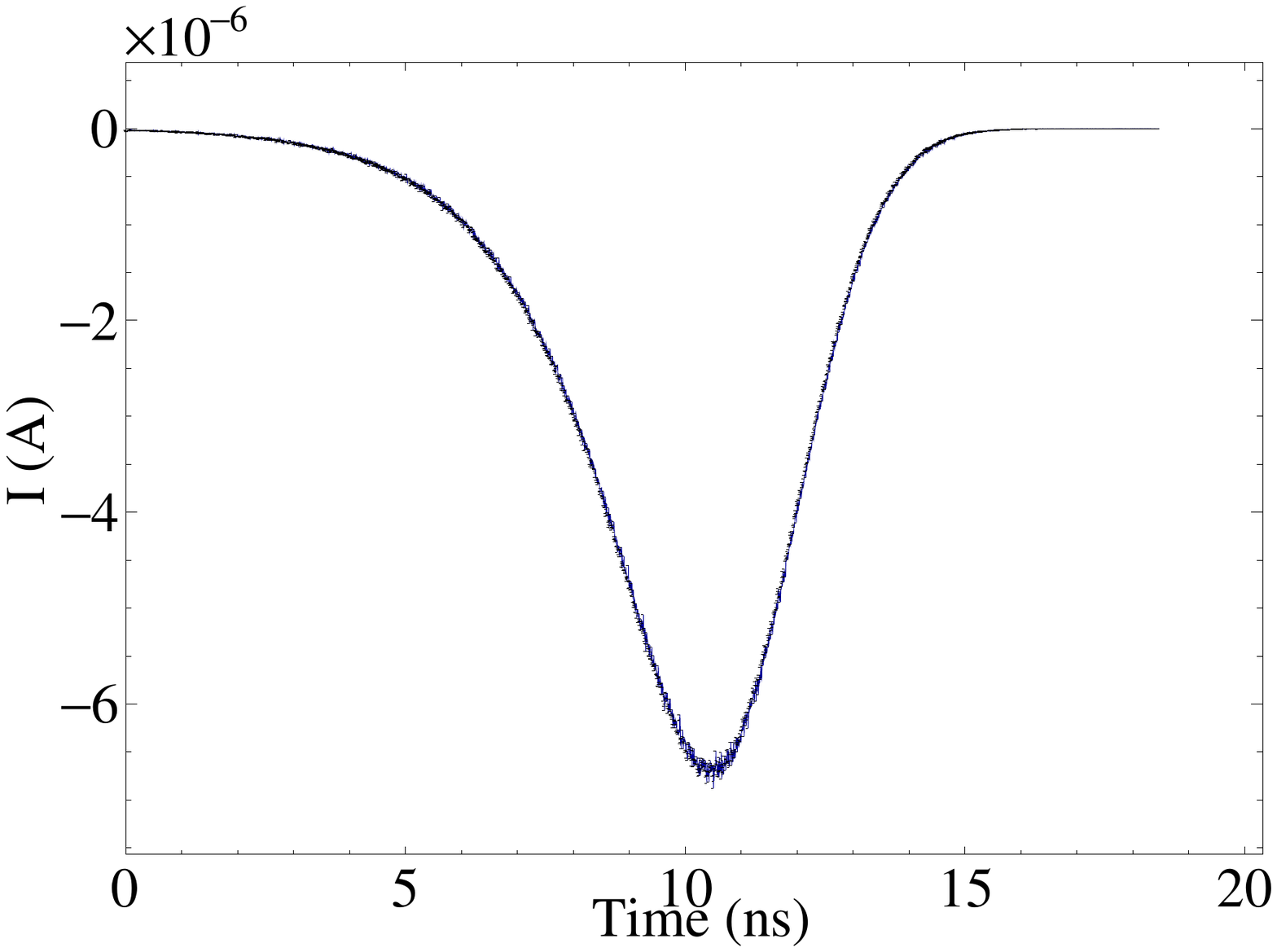}
    \caption{Example of the electric signal generated by the electrons on the bottom electrode in a parallel plate avalanche counter with a 1.6~mm gap.}
    \label{fig:ramo_el}
\end{figure}

As the different particle charges and types (electrons or ions) are known at any time during the simulation, the signals created by each type are generated separately.
Figure~\ref{fig:ramo_el} shows an example of the electric signal generated by the electrons on the bottom electrode in a parallel plate avalanche counter with a 1.6~mm gap.

%
%

\section{Validation and results}
\label{sec:Results}
%
%
\subsection{Swarm parameters validation}
\label{ssec:swarms}

During the simulation, \our allows to extract several parameters related to the particles propagation every $n\Delta t$ time intervals, where the integer $n$ and the time step $\Delta t$ are chosen by the user. Among these parameters, the position of the electrons and the ions is used to calculate the average position $\mathbf{r}_{c}(t)$ of the electron cloud or of the ion cloud as well as the associated standard deviations $\sigma_{\mathbf{r}_{c}}(t)$ in the three dimensions. These quantities are then used to extract the main swarm parameters~\cite{LViehland}: the drift velocity $V_d$, the transverse diffusion coefficient $D_{t}$ normalised by the mobility $\mu$ and $(\alpha-\eta)/N$ the first reduced Townsend coefficient corrected for the attachment rate $\eta$ and related to the gas gain. 

To compare the electron swarm parameters obtained with \our to literature data as well as online (LxCat) and offline (PyBoltz) values, a simple parallel plate avalanche counter was simulated using the \emph{static} field option.
It consisted of two 5~$\times$~5~cm$^2$ plates (one cathode and one anode) perpendicular to the $z$ axis and separated by a $1.6$~mm gap. The electric field was then oriented along the $z$ axis defined as the longitudinal axis while either the $x$ or $y$ axes correspond to the transverse direction. Swarm parameters were extracted from several simulations performed at different reduced electric fields E/N expressed in Townsend (1~Td = 10$^{-17}$V$\cdot$cm$^2$) and ranging from 10~Td to 1000~Td. We used iC$_4$H$_{10}$, CH$_4$, Ar/CO$_2$ (70/30)\% and Ar/CH$_4$ (70/30)\% as pure gases and mixtures.
The gas pressure was set to 1~bar and 10$^4$  primary particles were generated in a Gaussian ball (see figure~\ref{fig:GenPart}) of 10~\textmu m in diameter at 300~\textmu m from the cathode. The particles had then to drift through 1.3~mm of gas before being collected. For the gas gain measurement however, the pressure was reduced to 25~mbar and the number of primary electron/ion pairs was gradually decreased when increasing the electric field to maintain the number of produced particles in the avalanche below 10$^7$.
Results for each of the swarm parameters are presented in the following sections.

\subsubsection{Electron drift velocity}

The components of the electronic cloud velocity $\mathbf{v}_c(t)$ were calculated every $n\Delta t$ time steps using the electronic cloud average position $\mathbf{r}_c(t)$:

\begin{equation}
     \mathbf{v}_c(t) = \frac{\mathbf{r}_c(t) - \mathbf{r}_c(t-n\Delta t)}{n\Delta t}.
     \label{eq:vd}
\end{equation}

\begin{figure}[htb!]
    \begin{center}
        \includegraphics[width=.65\linewidth]{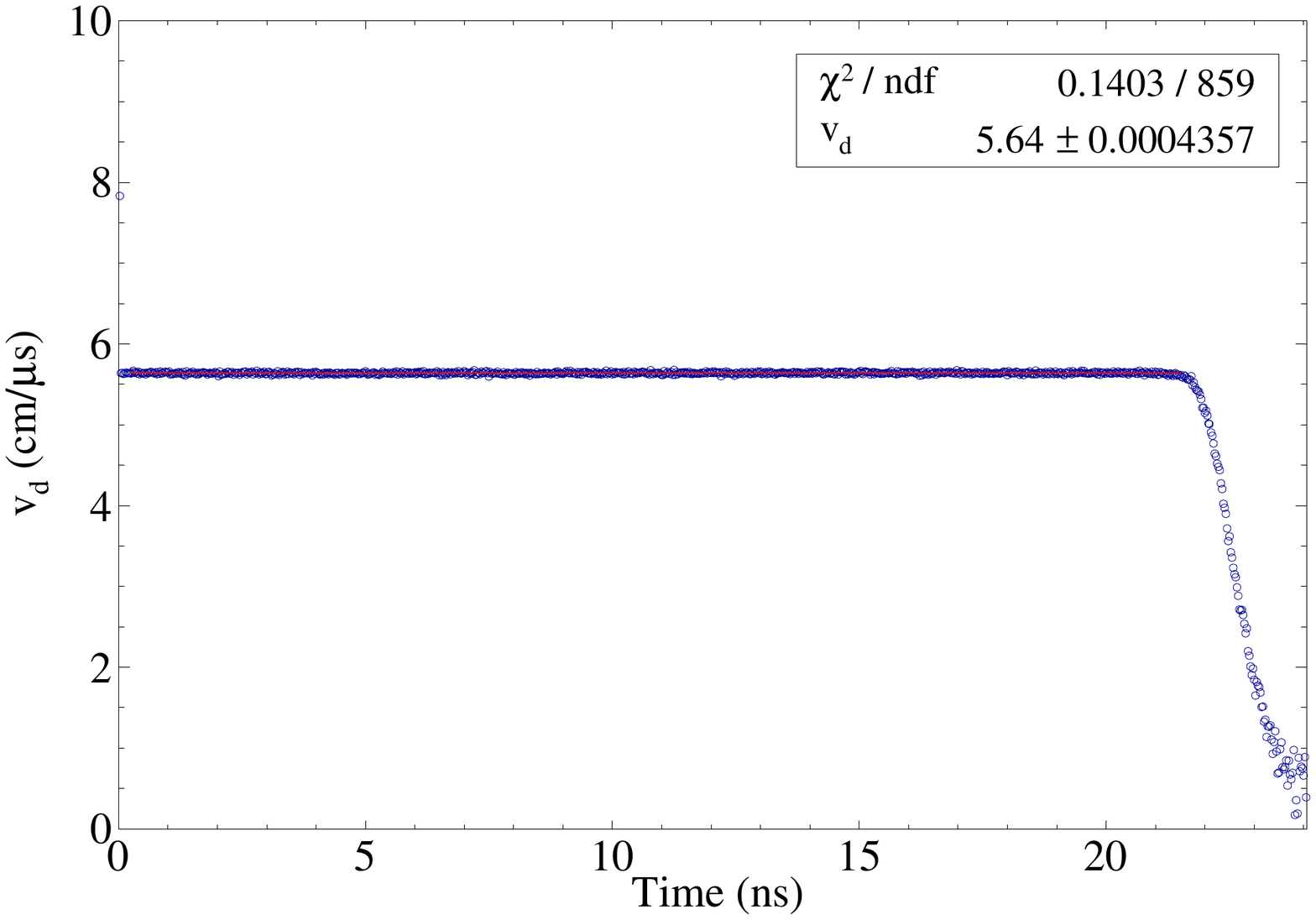}
    \end{center}
\caption{Electronic cloud velocity in the direction of the field as a function of time in isobutane at 100~Td along with the fit applied in the plateau region to extract the drift velocity (red line). This red line is indistinguishable from the data points. The electrons start being collected by the bottom electrode after a drift of about 22~ns.}
\label{fig:vd}
\end{figure}

The drift velocity $V_d$ corresponds to the projection of the cloud velocity $\mathbf{v}_c$ on the electric field vector and is extracted by fitting the time distribution with a constant value in the plateau region before electrons start being collected by the electrode. Figure~\ref{fig:vd} shows an example of the drift velocity for isobutane at a 100~Td reduced electric field as well as the fit applied to the plateau region with $n\Delta t=25$~ps.

\begin{figure}[htb!]
    \hbox{\hspace{-0.5cm} \includegraphics[width=.525\linewidth]{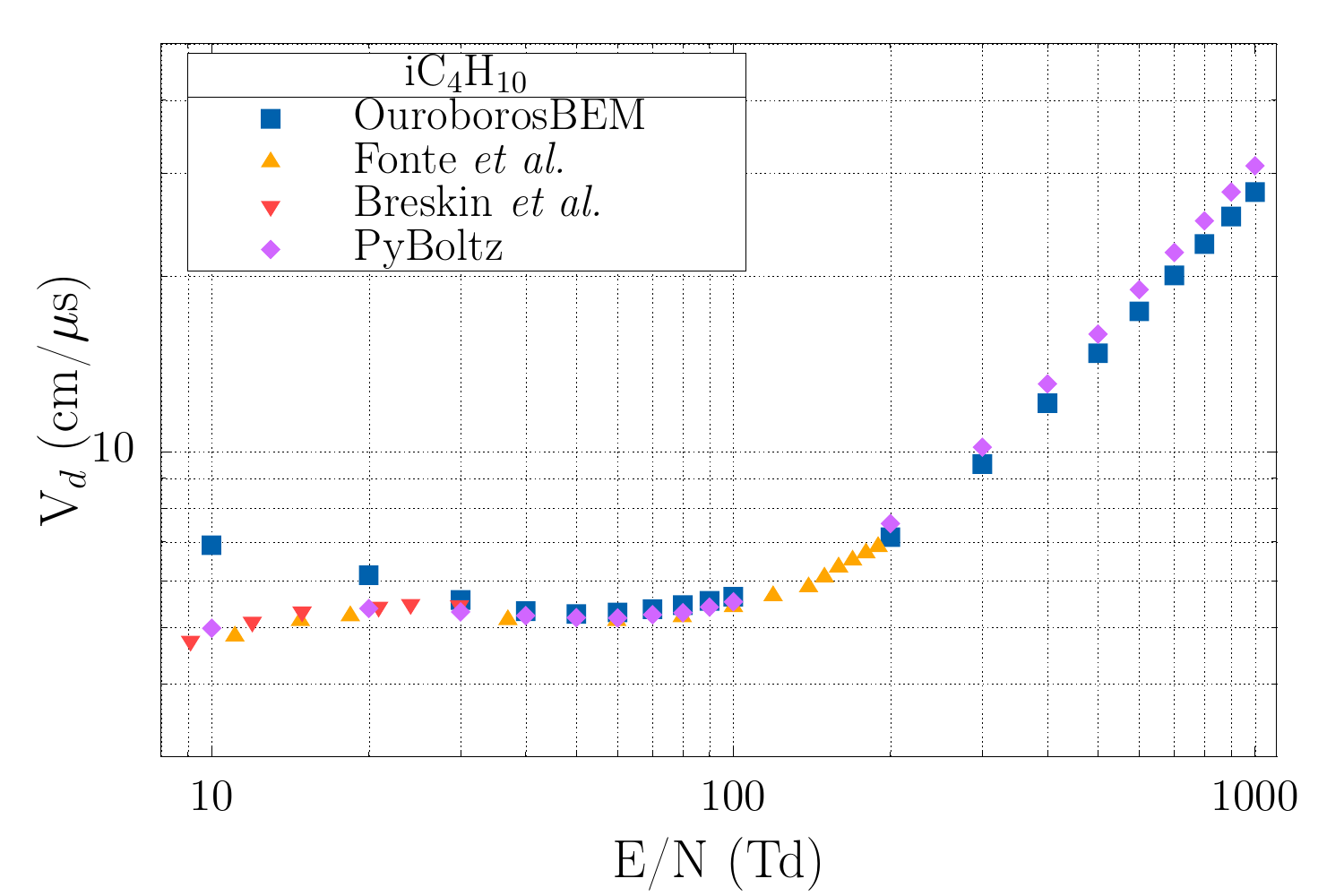}
          \hspace{-0.5cm} \includegraphics[width=.525\linewidth]{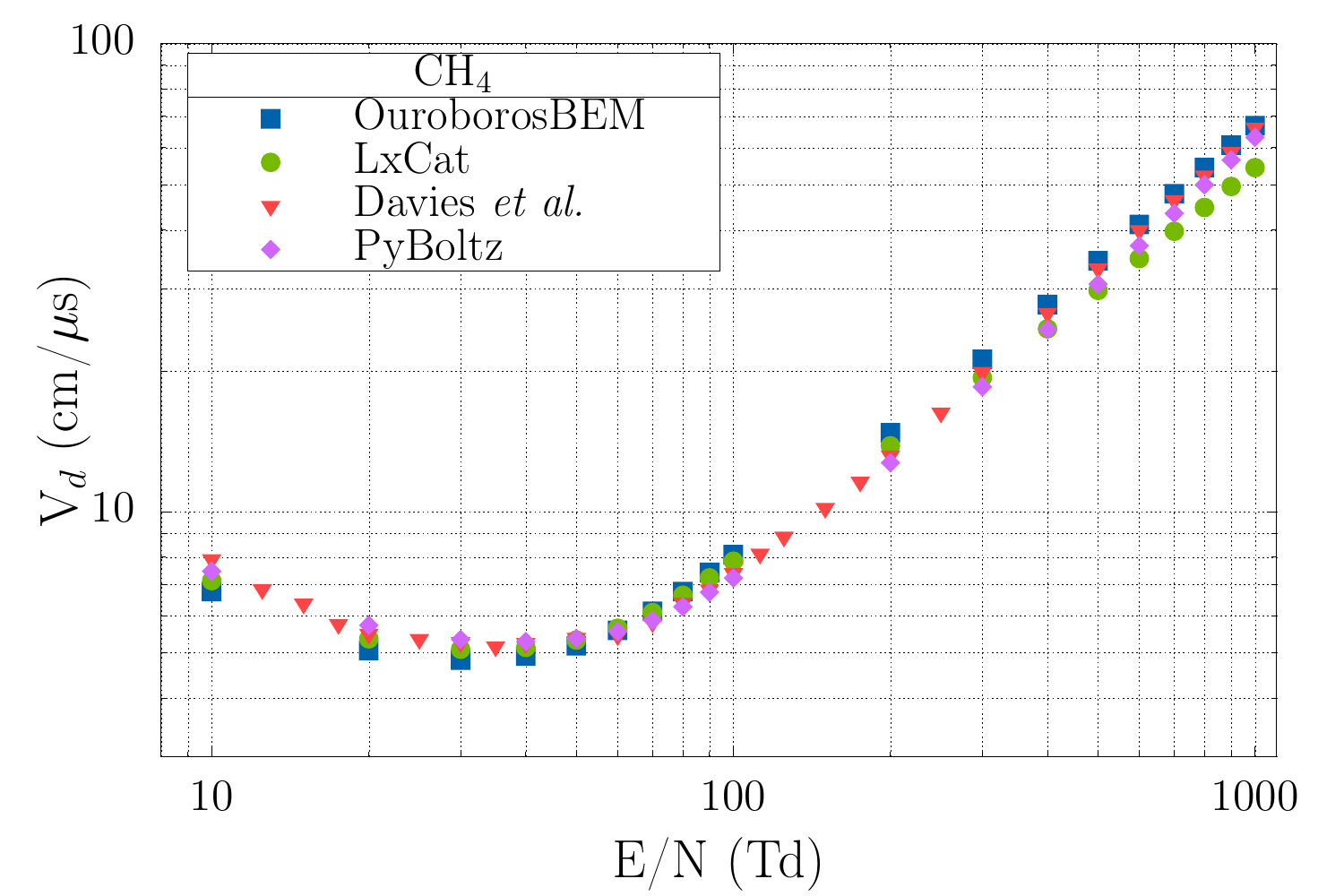}}
\caption{Drift velocities extracted for isobutane (left) and methane (right) at 1~bar as a function of the reduced electric field in Td (1~Td = 10$^{-17}$V$\cdot$cm$^2$). Data for LxCat were extracted online using the same dataset of the cross sections used in the software~\cite{CH4}. Data from Fonte et al. were extracted from~\cite{FONTE}, Breskin et al. from~\cite{BRESKIN} and Davies et al. from~\cite{Davies}.}
\label{fig:MeasuredVd}
\end{figure}

Figure~\ref{fig:MeasuredVd} shows the results obtained for pure isobutane and methane alongside literature data and online and offline calculations. The results show very good agreement except for values under 30~Td in the case of isobutane.

%
%
\subsubsection{Transverse diffusion coefficient}

The transverse diffusion coefficient $D_{t}$ was extracted by fitting the time distribution of the sigma value $\sigma_{t}$ of the spatial distribution of the electronic cloud in the transverse plane ($x$ axis in this example) with respect to the drift direction using the following function: 
\begin{equation}
     \sigma_{t}(t) = \sqrt{2D_{t}t}.
     \label{eq:diffusion}
\end{equation}

It was then normalized by the mobility extracted through the drift velocity following the relation $\mu=\frac{V_d}{E}$, with $E$, the norm of the electric field.

\begin{figure}[htb!]
    \begin{center}
        \includegraphics[width=.65\linewidth, clip=true, trim=0 0 0cm 0cm]{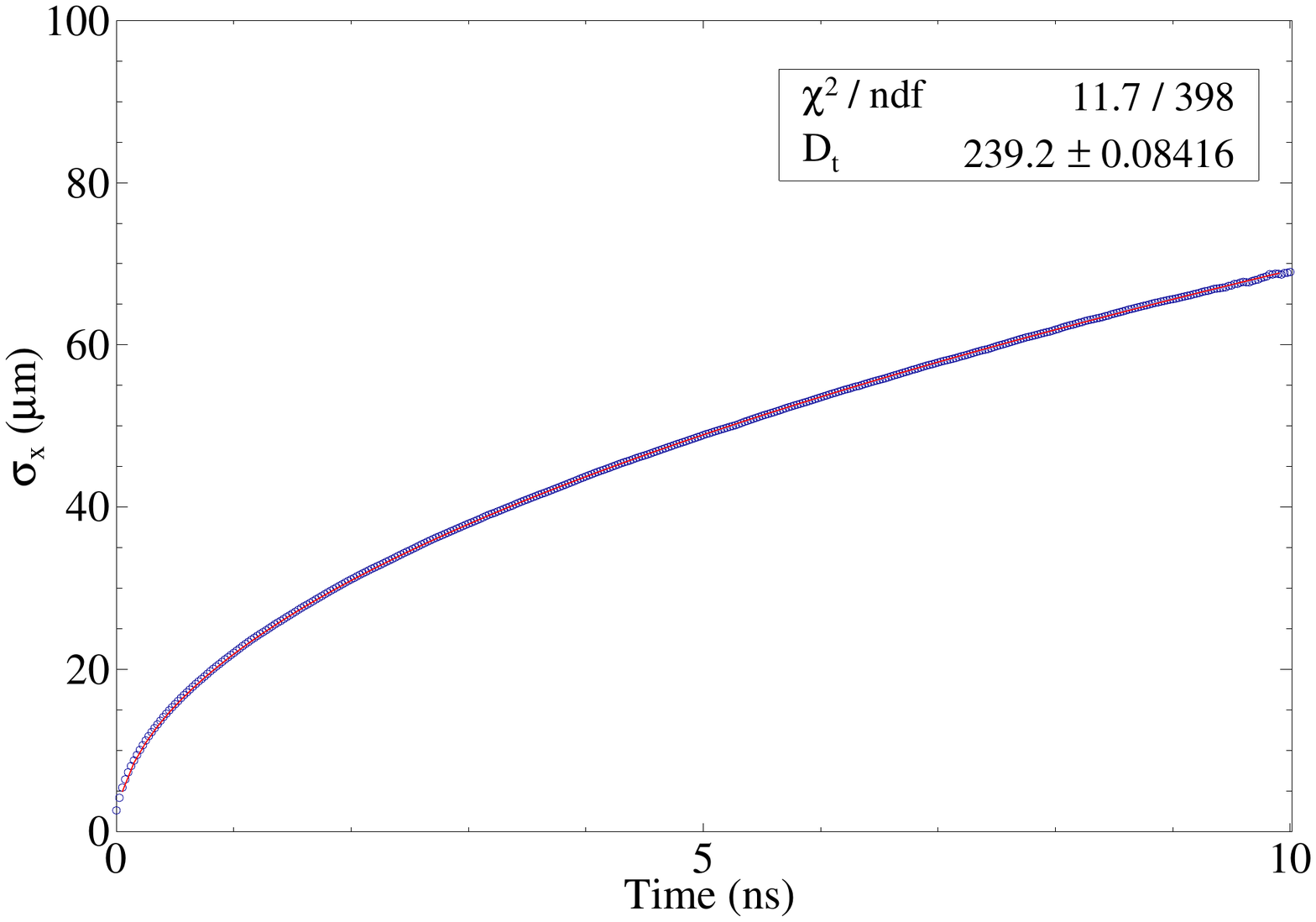}
    \end{center}
\caption{Sigma value in the $x$ axis of the electronic cloud spatial distribution as a function of time used for extracting the transverse diffusion coefficient $D_{t}$. The red line, which is indistinguishable from the data points, represents the fit applied using~\eqref{eq:diffusion}.}
\label{fig:Dt}
\end{figure}

Figure~\ref{fig:Dt} shows an example of the sigma value in the transverse plane (here in the $x$ axis) of the electronic cloud as a function of time in the case of a mixture of Ar/CO$_2$ (70/30)\% at 100~Td. The fit using~\eqref{eq:diffusion} applied to extract the diffusion coefficient is also represented as a solid red line.

\begin{figure}[htb!]
    \hbox{\hspace{-0.5cm} \includegraphics[width=.525\linewidth]{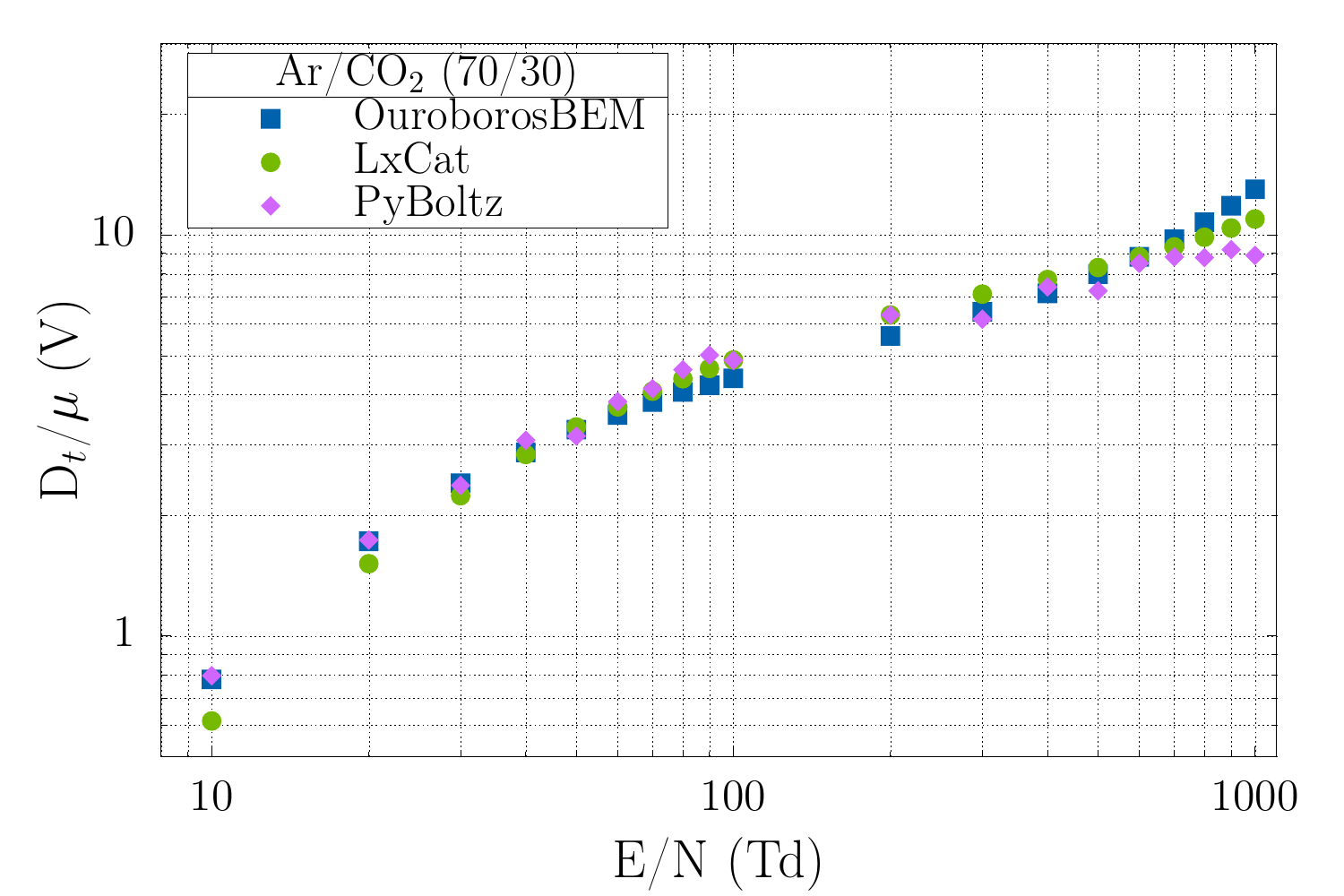}
          \hspace{-0.5cm} \includegraphics[width=.525\linewidth]{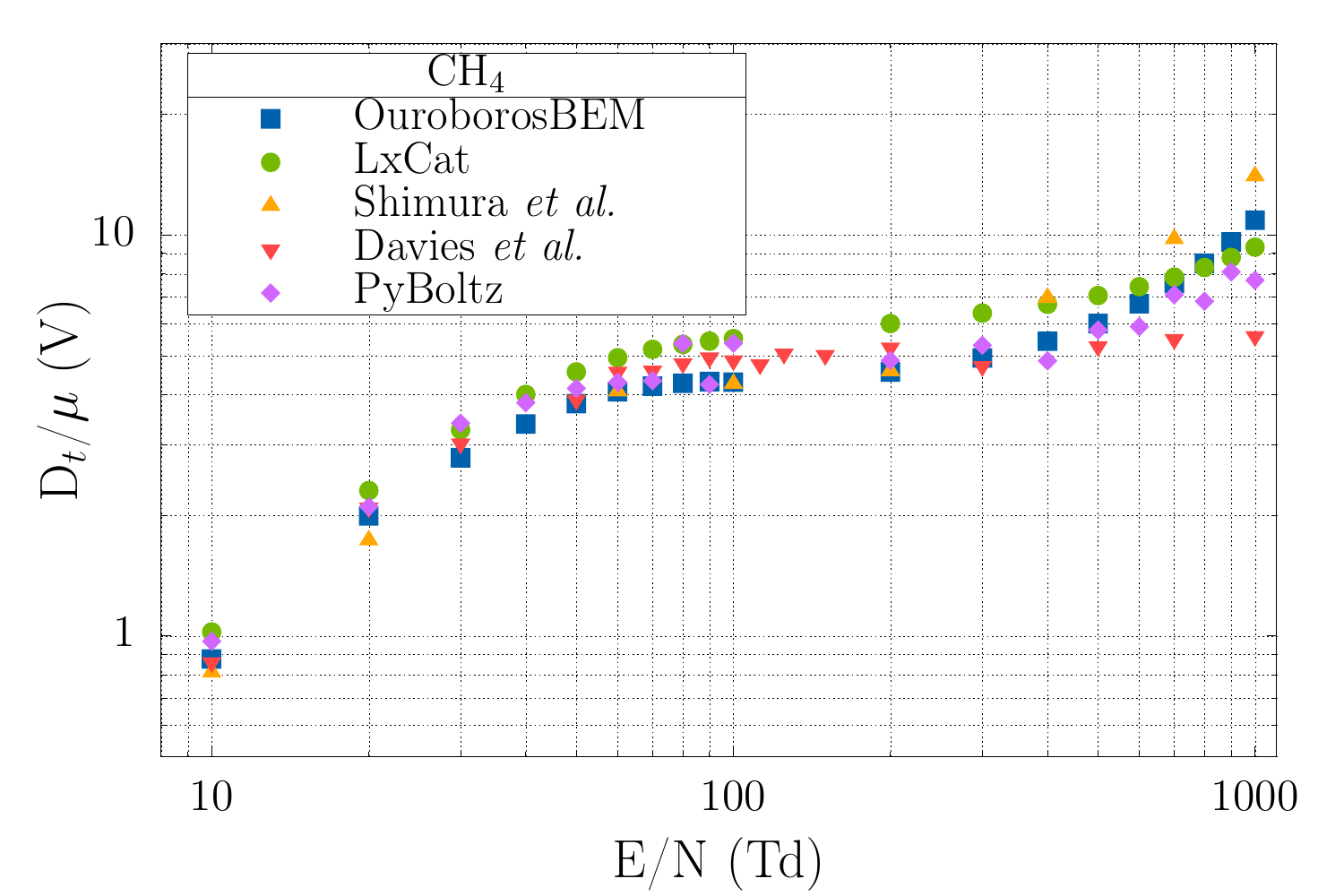}}
\caption{Transverse diffusion coefficient normalized by the mobility extracted for an Ar/CO$_2$ (70/30)\% mixture (left) and for pure methane (right) at 1 bar as a function of the reduced electric field in Td. Data for LxCat were extracted online using the same dataset of the cross sections used in the software~\cite{magboltz,CO2,CH4}. Data for Shimura et al. were extracted from~\cite{Shimura_1992}. Data for Davies et al. were extracted from~\cite{Davies} and are based on Monte Carlo predictions.}
\label{fig:MeasuredDt}
\end{figure}

Figure~\ref{fig:MeasuredDt} shows the results obtained for the Ar/CO$_2$ (70/30)\% mixture and for pure methane alongside online and offline calculations and literature data. The results show slight discrepancies in the case of Ar/CO$_2$ especially at high reduced electric field, i.e.~above 700~Td, whereas the agreement for methane is more mitigated. However in this case, a strong consensus cannot be seen among the literature data and the results seem to fit with the data from Shimura et al.~\cite{Shimura_1992} under 200~Td. The behaviour is in any case well reproduced overall.

\subsubsection{First Townsend coefficient}

The first Townsend coefficient $\alpha$ relies directly to the gas gain of the avalanche process. It can be extracted using the gain $G$ measured in the simulation:

\begin{equation}
     G = e^{\alpha d},
     \label{eq:townsend}
\end{equation}
which corresponds to the number of electrons collected on the detector anode and produced through a drift distance $d$ divided by the number of primary electrons. As the attachment processes can occur during the transport of the electrons through the detector gap, the results are presented as the reduced first Townsend coefficient corrected by the attachment process $\eta$.

\begin{figure}[htb!]
    \hbox{\hspace{-0.5cm} \includegraphics[width=.525\linewidth]{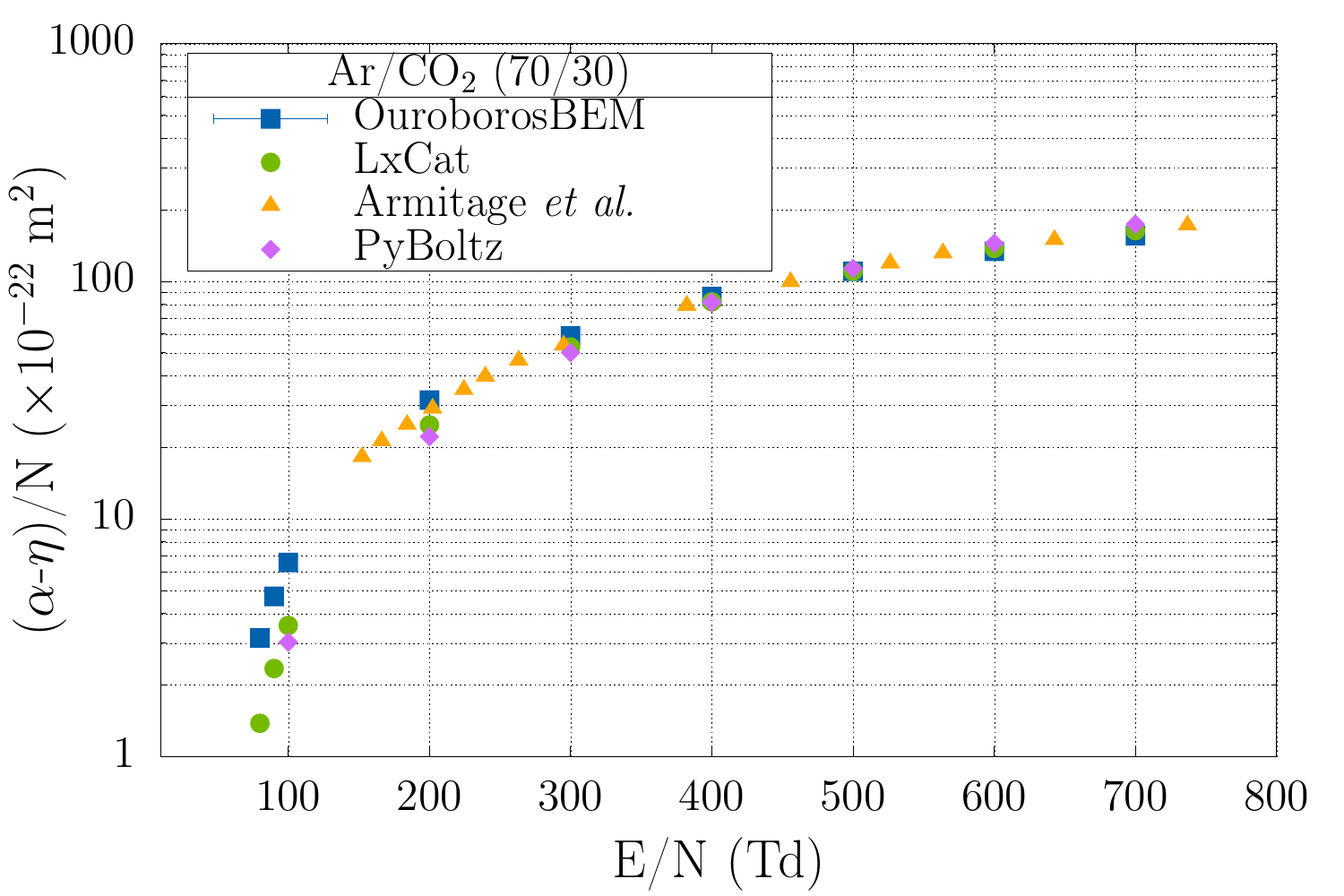}
          \hspace{-0.5cm} \includegraphics[width=.525\linewidth]{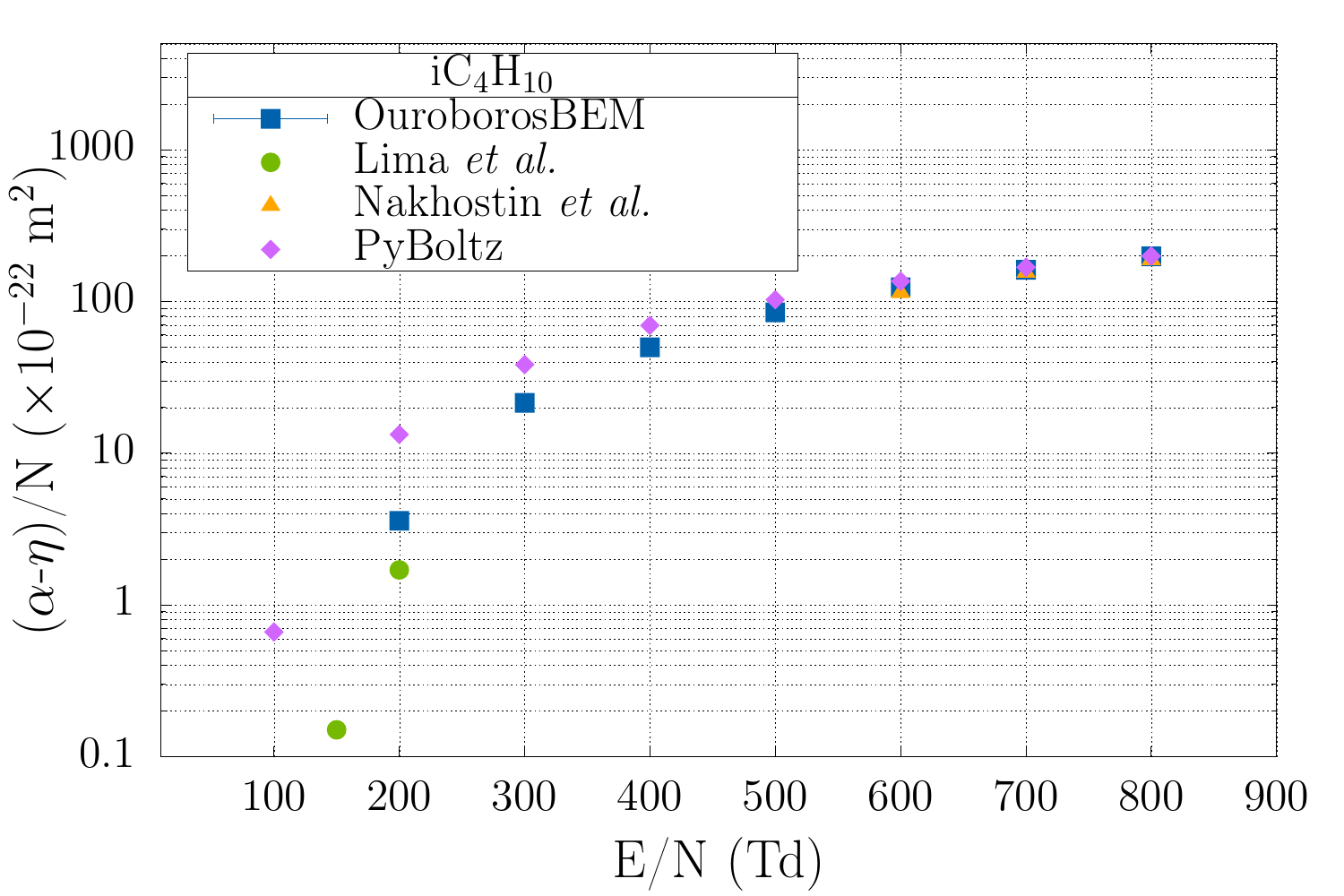}}
\caption{Reduced first townsend coefficient $\alpha$ corrected for the attachment coefficient $\eta$ for an Ar/CO$_2$ (70/30)\% mixture (left) and pure isobutane (right) at 25~mbar as a function of the reduced electric field in Td. Data for LxCat were extracted online using the same dataset of the cross sections used in the software~\cite{magboltz,CO2}. Data for Armitage et al. were extracted from~\cite{ARMITAGE}. Data for Lima et al. were extracted from~\cite{LIMA} and data for Nakhostin et al. from~\cite{NAKHOSTIN}.}
\label{fig:alpha}
\end{figure}

Figure~\ref{fig:alpha} presents the results obtained for the Ar/CO$_2$ (70/30)\% mixture and for pure isobutane at 25~mbar as a function of the reduced electric field in Td.

Although the results are in very good agreement in the case of the Ar/CO$_2$ (70/30)\% mixture compared to literature data, some discrepancies can be seen for pure isobutane below 600~Td. The values obtained from the PyBoltz solver show larger differences compared to the experimental data from Lima~\textit{et al.}~\cite{LIMA} (between 100 and 200~Td) up to a an order of magnitude where the differences with \our are around a factor 2. It is however hard to conclude as the literature data in this range of reduced electric fields are scarce.

%
%
\subsection{Single and double GEM effective gain}

\begin{figure}[htb!]
    \centering
    \includegraphics[width=0.7\linewidth]{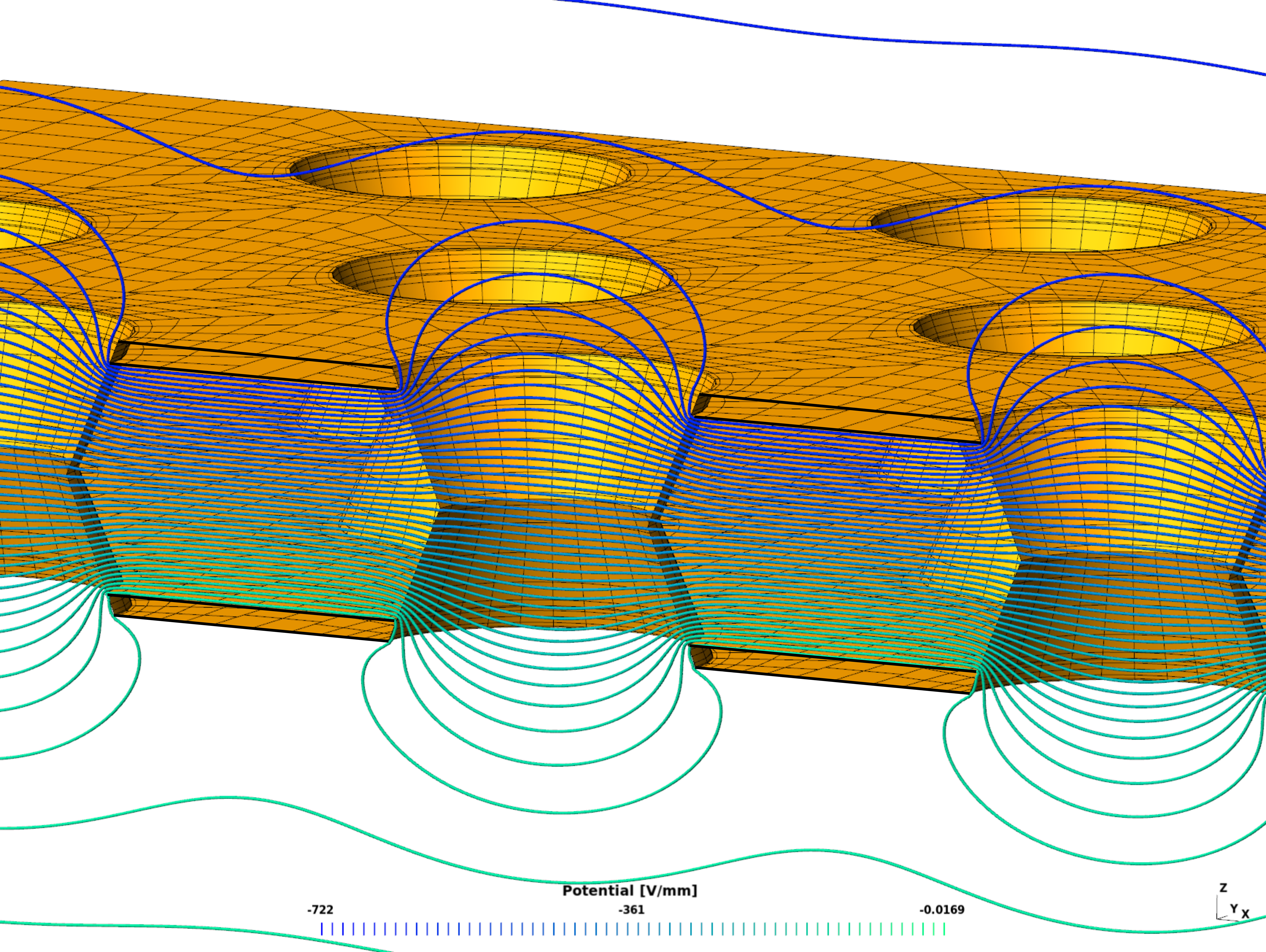}
    \caption{{\textsc{gmsh}} cross section view of a single GEM detector along with the iso-potential lines. Due to the GEM dimensions, the top and bottom electrodes are out of view here. The dielectric media is sandwiched between two electrodes and drilled with bi-conic holes.}
    \label{fig:GEM_mesh_and_pot}
\end{figure}

The effective gain measured in a single and a double GEM detector were measured for several differences of potential on the GEM structure $\Delta V_{\mathrm{GEM}}$. The GEMs used were standard designs from CERN where the bi-conic hole inner and outer diameters are 50 and 70~\textmu m with a pitch of 140~\textmu m. The drift field was set to 3~kV$\cdot$cm$^{-1}$ and the induction field to 4.4~kV$\cdot$cm$^{-1}$. In the case of the double GEM detector, the transfer field between the two meshes was set equal to the induction field. The gas mixture used was Ar/CO$_2$ (70/30)\% at atmospheric pressure. For the double GEM detector, both GEM structures have the same $\Delta V_{\mathrm{GEM}}$. Figure~\ref{fig:GEM_mesh_and_pot} shows \gmsh cross sections of the meshes generated for the single GEM geometry along with the iso-potential lines. The electric fields have been calculated using the \emph{static} field option of the simulation.
The effective gain was measured as the number of electrons collected at the anode divided by the number of primary electrons.

\begin{figure}[htb!]
    \begin{center}
        \includegraphics[width=.7\linewidth]{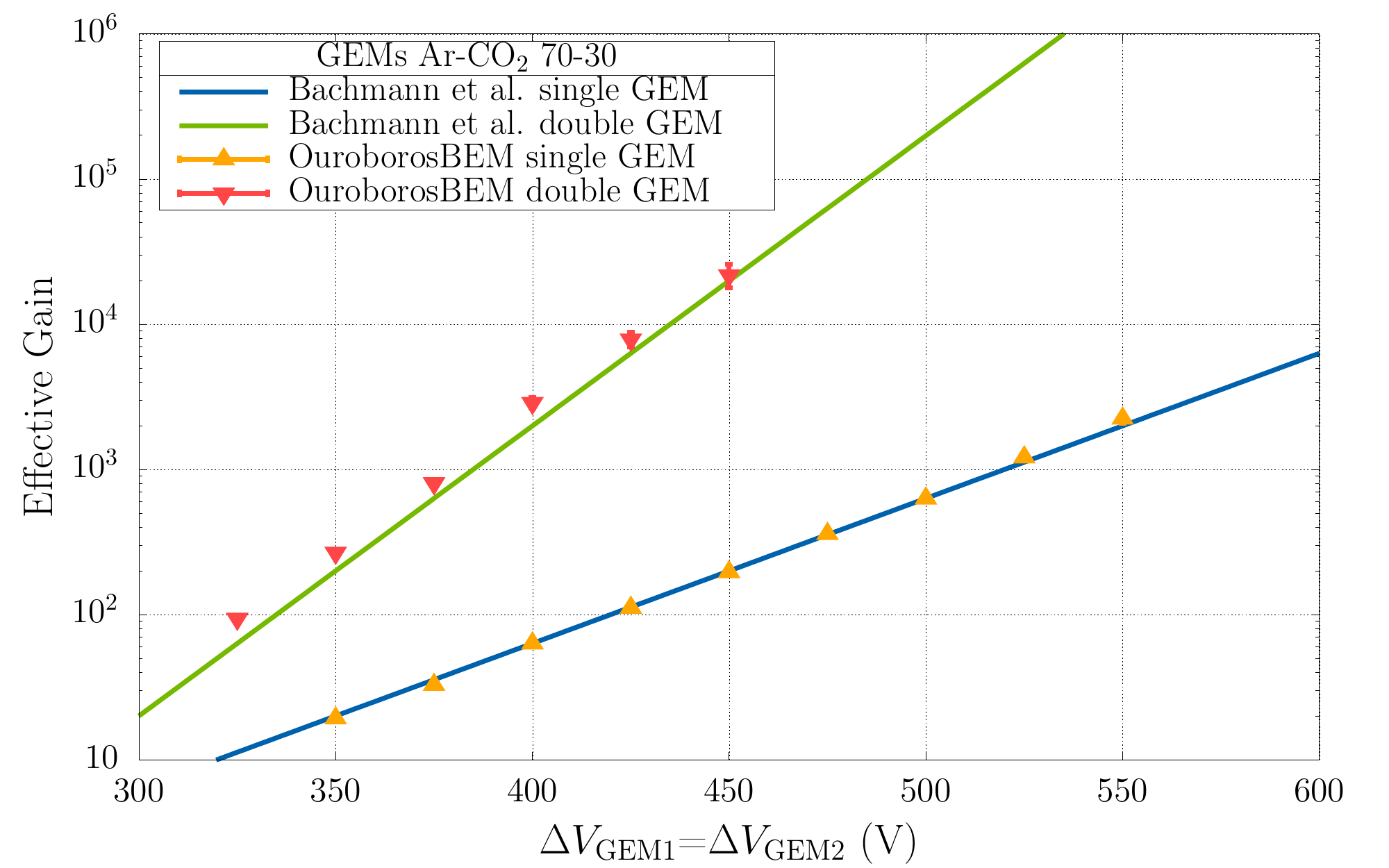}
    \end{center}
\caption{Comparisons of the effective gain measured in \our for a single and a double GEM detector in Ar/CO$_2$ (70/30)\% with results obtained experimentally by~\citep{BACHMANN}. The difference of potential in the double GEM configuration is equal for both structures.}
\label{fig:GainGEM}
\end{figure}

Figure~\ref{fig:GainGEM} shows the results obtained for the single and double GEM detectors compared to the data from Bachmann~{\it et al.}~\citep{BACHMANN}. The agreement for the single GEM detector is almost perfect while a small overestimation of the gain can be observed for the double GEM detector. In our case, the transfer region between the two detectors has been reduced to 1~mm instead of 3~mm to reduce the geometry size. This can explain the slight increase in the gain by an increase in the collection efficiency of the second detector. Each simulated configuration which microscopically generated and transported more than 2$\cdot$10$^6$ particles, took less than 20~min.

%
%
\subsection{Space charge effects in a single GEM detector}
\label{ssec:SpaceChargeResults}

The previous single GEM set-up was also simulated with all the dynamic effects enabled (influence, charging-up and N-body). The following figures illustrate the different space charge effects that may affect the GEM behaviour. The results are presented at two different simulated times, i.e. 6.1~ns and approximately 1.5~\textmu s where in that later case, all electrons have already been collected and only ions (cations and anions from the attachment processes) remain. The results are shown only for qualitative purposes and illustration of the software capabilities.

\begin{figure}[htb!]
    \begin{subfigure}[b]{1.0\textwidth}
        \caption{}
        \centering
        \includegraphics[width=.8\linewidth]{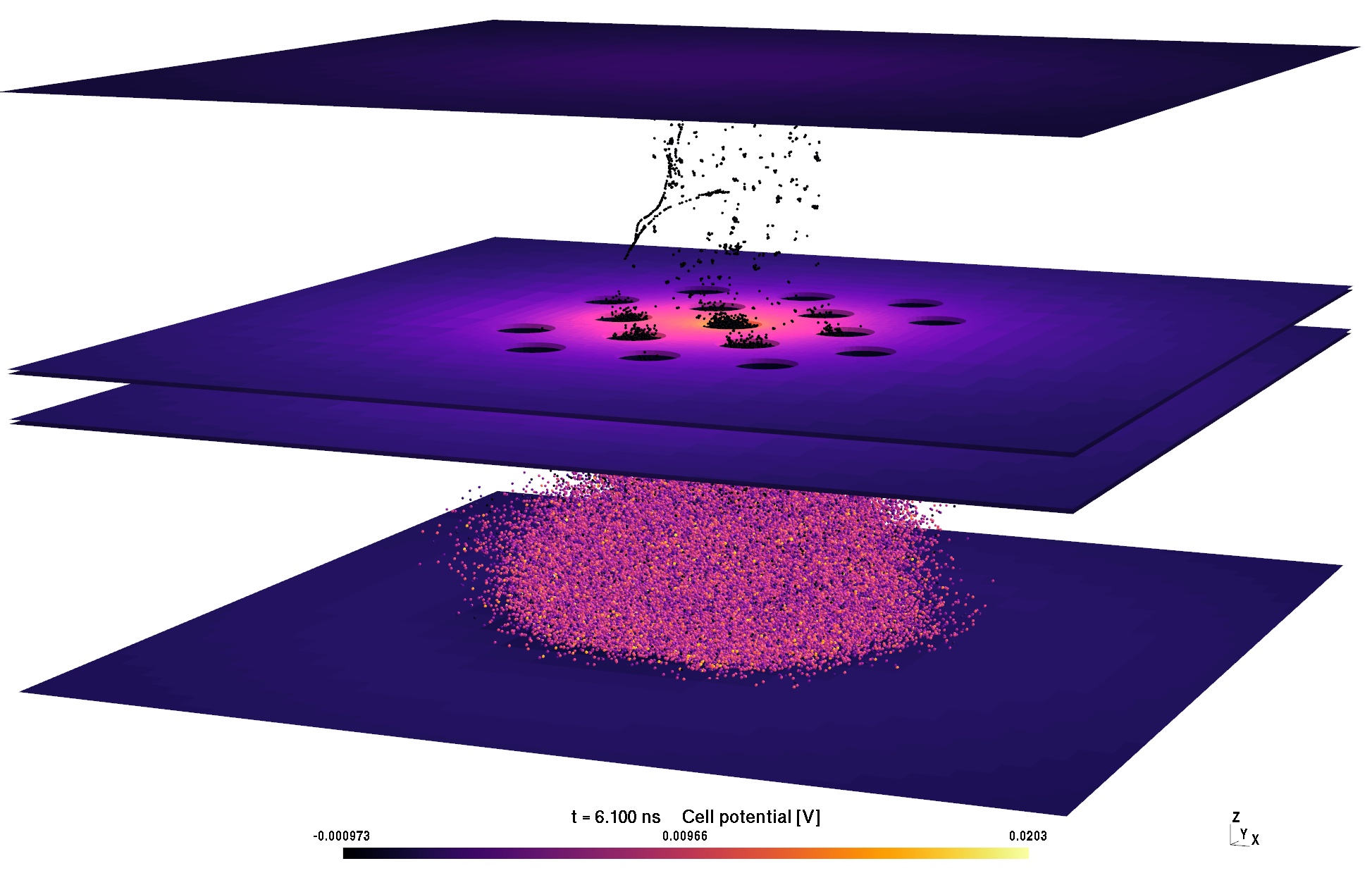}
    \end{subfigure}\\
     \begin{subfigure}[b]{1.0\textwidth}
        \caption{}
        \centering
        \includegraphics[width=.8\linewidth]{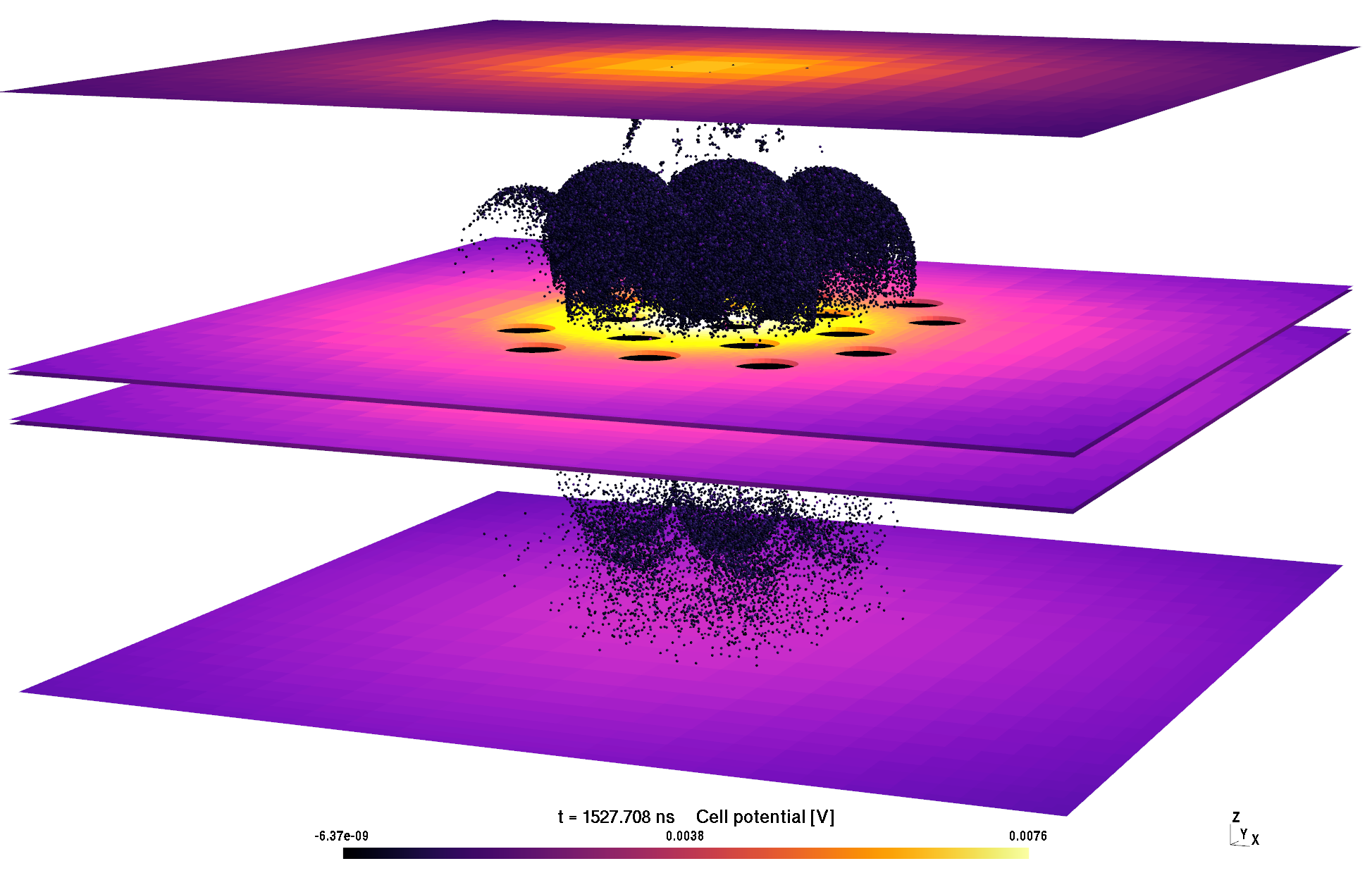}
    \end{subfigure}
    \caption{Influence of the particles (expressed in V) on the electrode cells of the GEM set-up at (a) 6.1~ns and (b) approximately 1.5~\textmu s. The particle positions have been added and their velocity is represented as colored dots. In view (a), most of the ions are still located inside the bi-conic holes whereas the much faster electrons have drifted away from the GEM structure and start being collected by the bottom electrode. In view (b), all the electrons have been collected and the slow anions and cations are drifting towards their corresponding electrodes. Note that the color scale is different for each picture.}
    \label{fig:Inlfuence}
\end{figure}

\begin{figure}[htb!]
    \centering
    \includegraphics[width=.75\linewidth]{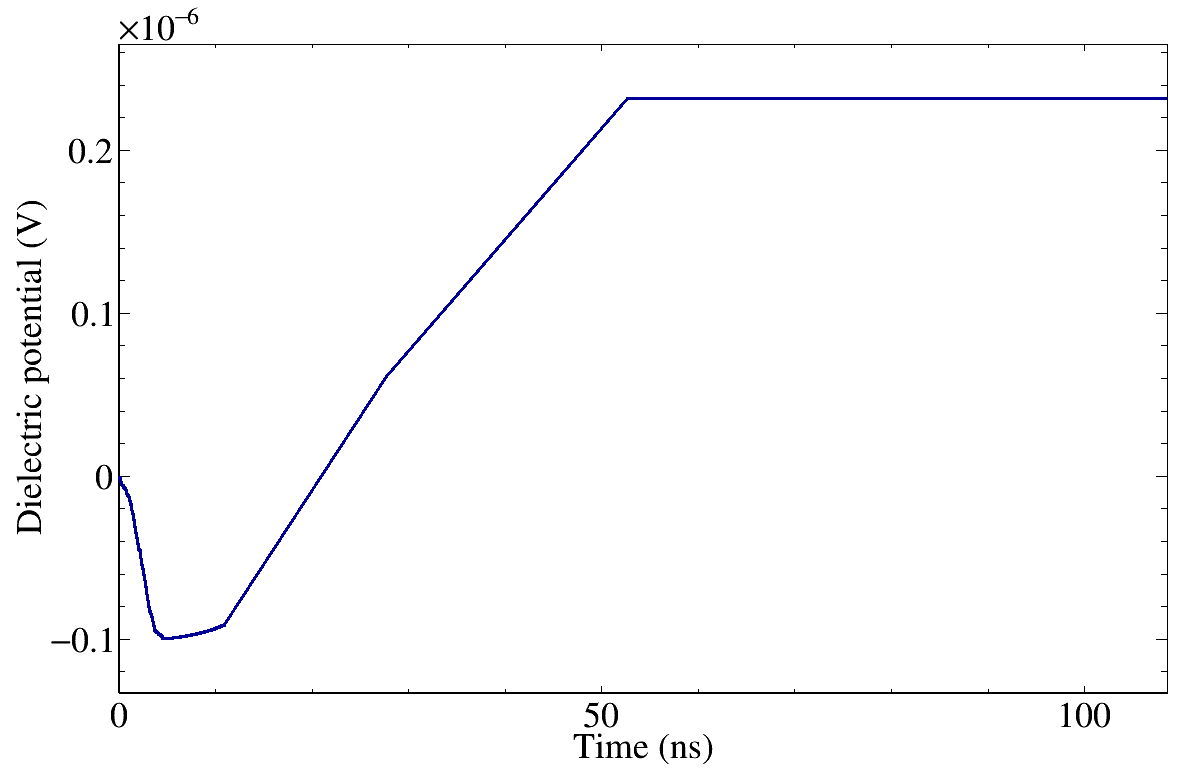}
    \caption{Potential in V created on the dielectric as a result of the charging-up process as a function of time. The potentials on the dielectric cells making the bi-conic holes have been summed.}
    \label{fig:chargingUp}
\end{figure}

\begin{figure}[htb!]
    \begin{subfigure}[b]{0.48\textwidth}
        \caption{}
        \centering
        \includegraphics[width=0.95\linewidth]{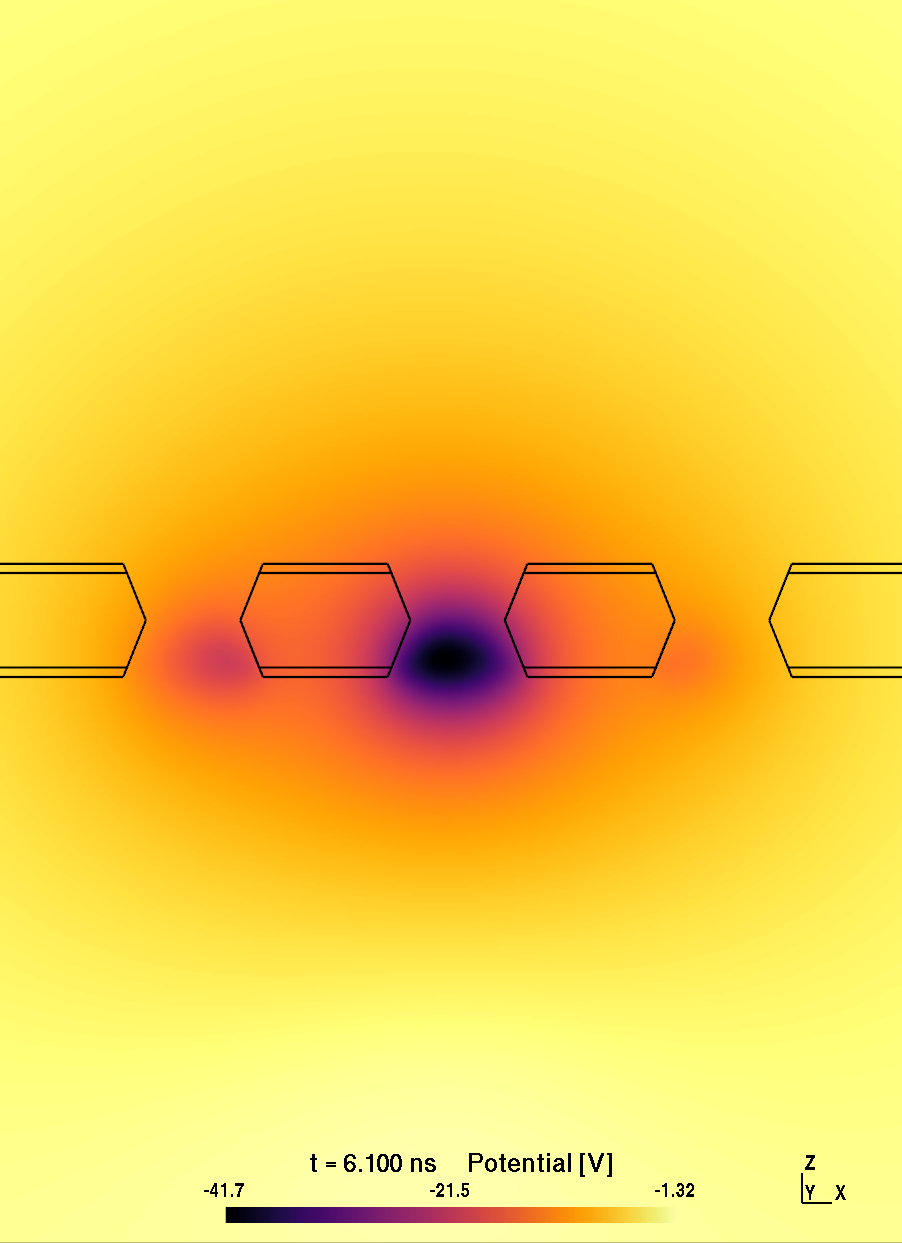}
    \end{subfigure}
    \hfill
     \begin{subfigure}[b]{0.48\textwidth}
        \caption{}
        \centering
        \includegraphics[width=0.95\linewidth]{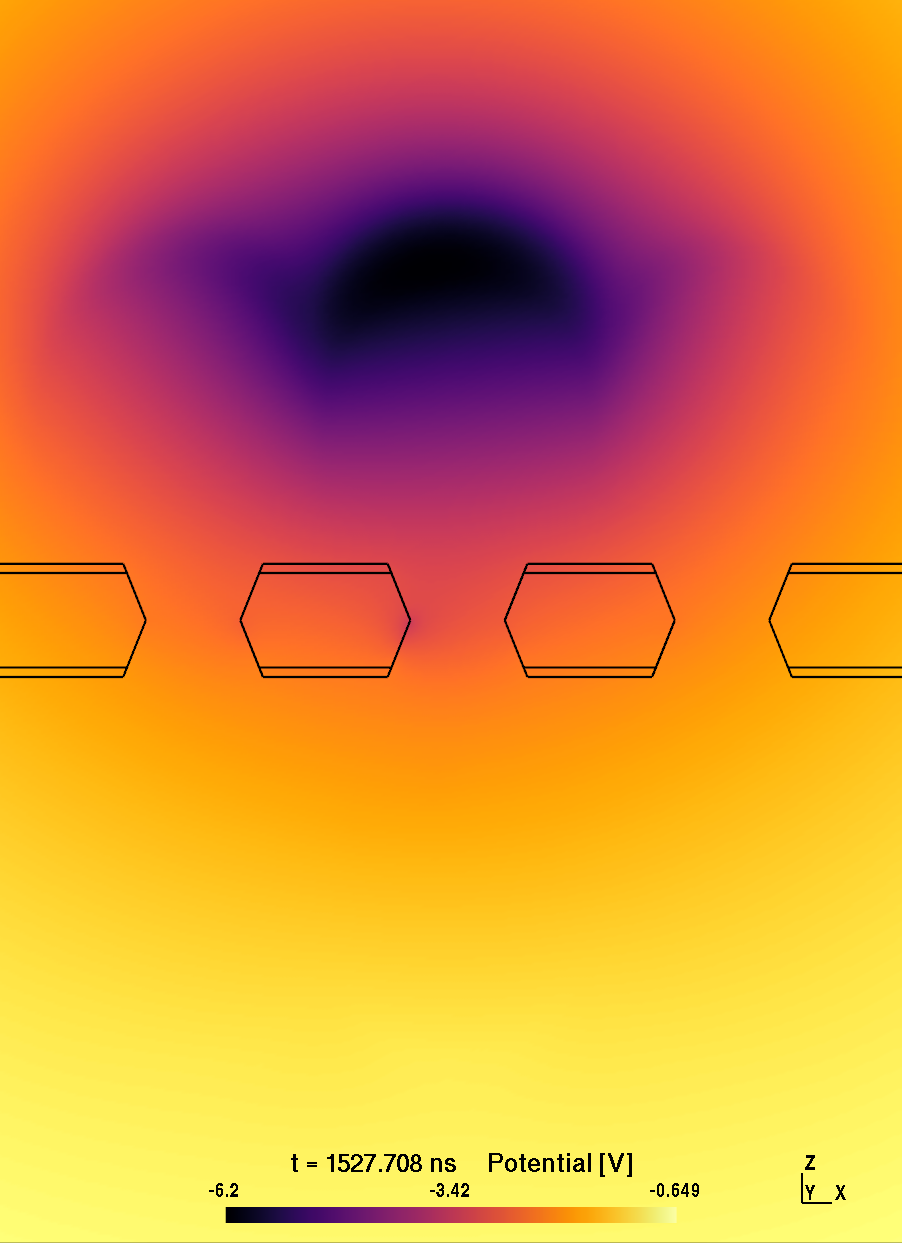}
    \end{subfigure}
    \caption{Effects of the space charge in the single GEM set-up on the electric potential at (a) 6.1~ns and (b) approximately 1.5~\textmu s. The potential generated by the GEM itself was subtracted from the results. Note that the color scale is different for each picture.}
    \label{fig:SpaceCharge}
 \end{figure}

Figure~\ref{fig:Inlfuence} shows the influence of the particles on the different cell surfaces expressed in volts. The potentials applied on the electrodes have been subtracted from the results.
Figure~\ref{fig:chargingUp} represents the charging-up of the dielectric materials expressed in volts as a function of time. The first 5~ns correspond to fast electrons being attached on the dielectric bi-conic holes. After 5 ns, most of the electron cloud has drifted away from the GEM structure and the much slower ions start being captured and therefore slowly compensate the overall negative potential. The fact that the curve looks broken after approximately 12~ns is due to the increase in the time step imposed to track ions more rapidly once all the electrons have been collected on the bottom electrode. The cumulative effect of all the space charge processes on the detector electric potentials are presented in Figure~\ref{fig:SpaceCharge}. The electric potential of the bare GEM set-up was also subtracted from the results.


%
%

\clearpage
\section{Conclusion}
\label{sec:Conclusion}

In this work we have presented the \our software dedicated to the simulation of gaseous detectors.
This program, written in CUDA/C++ language, runs on multi-GPU systems equipped with nVidia devices.
A BEM solver including geometry description and meshing has been implemented to solve the electrostatic problem for configurations with both electrodes and dielectric media. When selecting the \emph{dynamic} field option, this solver is called at each time step to compute the electric field generated by the set-up and includes several space charge effects. Those consist as the influence of a large number of charged particles on the overall field and the charging-up effect on the dielectric materials. \our also computes the Coulombian repulsion or attraction between all particles. This N-body problem can be treated in reasonable computing times for up to a few $10^7$ particles thanks to the power of the GPU devices. 
The software handles different gas species (pure gas or admixtures) in which are created, tracked and cascaded some electron-ion pairs. Last but not least, \our allows to generate the pulse shape of the signal measured on defined read-out electrodes.
With the determination of swarm parameters from parallel plate detectors and of the gas gain of simple and double GEM detectors, \our has clearly demonstrated its capacities to accurately simulate the physical processes involved in gaseous detectors and therefore appears as a strong challenger to the famous \garfield.

\our is not limited to the simulations of gaseous detectors and has already been used to simulate some radio-frequency Paul traps where radioactive ions are trapped within a cooling buffer gas. Other applications could be handled, without changing the general workflow of the program, by adding the proper interaction cross sections and processes as for instance in liquid noble gases such as argon used in time projection chambers (LArTPC) for some neutrino physics experiments~\cite{abi2020deep} or xenon in dark matter searches~\cite{XENON1T}.

For the future, several enhancements of the software are envisaged. One of them consists in the implementation of the fast multipole method (FMM) in order to improve the precision on the forces due to the N-body Coulombian interactions. Another useful enhancement requires modifying the geometry description and meshing phases in order to ease the interplay with CAD software used in the design of the experimental set-ups.
%
%
\acknowledgments
The authors would like to thank the R\'egion Normandie via its R\'eseaux d'Int\'er\^ets Normands (grant RIN-THESMOG 18E01634/18P02356) for financially supporting the acquisition of a GPU server. Part of the simulations presented were also performed using the computing resources of the Centre R\'egional Informatique
et d'Applications Num\'eriques de Normandie (CRIANN, Normandy, France)

%
%
\bibliography{OuroborosBEM}

\end{document}